\documentclass[10pt,journal,final,twocolumn,]{IEEEtran}

\usepackage{CJK}
\usepackage{color, soul, framed}
\usepackage{algorithm}
\usepackage{algorithmic}
\usepackage{amsmath}
\usepackage{amsthm}
\usepackage{amssymb}
\usepackage{psfrag}
\usepackage{graphicx}
\usepackage{epstopdf}
\usepackage{multirow}
\usepackage{stfloats}
\usepackage{cite}
\usepackage{subcaption}
\usepackage{color}
\usepackage[normalem]{ulem}
\usepackage{arydshln}
\usepackage{textcomp}
\usepackage{enumerate}
\usepackage{bbm}
\usepackage{xcolor}
\usepackage{xpatch}
\usepackage{array}
\usepackage{makecell}
\usepackage{multirow}
\usepackage{suffix}
\usepackage{mathtools}
\usepackage{diagbox}
\allowdisplaybreaks[4]

\newtheorem{lemma}{\bf{Lemma}}
\newtheorem{proposition}{\bf{Proposition}}

\newtheorem{remark}{\bf{Remark}}
\title{Multi-User Multi-IoT-Device Symbiotic Radio:\\ A Novel Massive Access Scheme \\ for Cellular IoT}

\author{Jun Wang, Ying-Chang Liang, \emph{Fellow, IEEE}, and Sumei Sun, \emph{Fellow, IEEE}
\thanks{
Part of this work was presented in IEEE ICC 2023\cite{wang2023icc}. 

J. Wang is with the National Key Laboratory of Science and Technology on Communications, University of Electronic Science and Technology of China (UESTC), Chengdu 611731, China (e-mail: junwang@std.uestc.edu.cn). 
Y.-C. Liang is with the Center for Intelligent Networking and Communications, University of Electronic Science and Technology of China, Chengdu 611731, China (e-mail: liangyc@ieee.org). 
S. Sun is with the Institute for Infocomm Research, Agency for Science, Technology and Research, Singapore 138632 (e-mail: sunsm@i2r.a-star.edu.sg).
}
}
\begin{document}
 \maketitle
\begin{abstract}
Symbiotic radio (SR) is a promising technique to support cellular Internet-of-Things (IoT) by forming a mutualistic relationship between IoT and cellular transmissions. In this paper, we propose a novel multi-user multi-IoT-device SR system to enable massive access in cellular IoT. In the considered system, the base station (BS) transmits information to multiple cellular users, and a number of IoT devices simultaneously backscatter their information to these users via the cellular signal. The cellular users jointly decode the information from the BS and IoT devices. Noting that the reflective links from the IoT devices can be regarded as the channel uncertainty of the direct links, we apply the robust design method to design the beamforming vectors at the BS.
Specifically, the transmit power is minimized under the cellular transmission outage probability constraints and IoT transmission sum rate constraints. The algorithm based on semi-definite programming and difference-of-convex programming is proposed to solve the power minimization problem. Moreover, we consider a special case where each cellular user is associated with several adjacent IoT devices and propose a direction of arrival (DoA)-based transmit beamforming design approach. The DoA-based approach requires only the DoA and angular spread (AS) of the direct links instead of the instantaneous channel state information (CSI) of the reflective link channels, leading to a significant reduction in the channel feedback overhead. Simulation results have substantiated the multi-user multi-IoT-device SR system and the effectiveness of the proposed beamforming approaches. It is shown that the DoA-based beamforming approach achieves comparable performance as the CSI-based approach in the special case when the ASs are small.


\end{abstract}

\section{Introduction}
The Internet of Things (IoT) has experienced remarkable growth in recent years. According to the latest IoT report released in March 2023, the number of global IoT connections has reached 15.14 billion in 2023 and is projected to surpass 25 billion within the coming seven years~\cite{IoTreport}. The IoT has impressively empowered a variety of newly emerging services and applications, such as smart home, intelligent transportation, and smart manufacturing, significantly enhancing our daily lives~\cite{you2021towards}. 
Cellular IoT has garnered substantial attention from both academia and industry to support these massive IoT devices~\cite{sharma2019toward,benhiba2018comparative}. Rather than constructing dedicated networks and infrastructure for IoT transmission, cellular IoT leverages existing cellular networks to facilitate IoT connections. The Long Term Evolution (LTE) for Machine-Type Communications (LTE-M) and Narrow-Band IoT (NB-IoT), developed in the 3rd Generation Partnership Project (3GPP), are primary cellular IoT standards and have been extensively deployed~\cite{liberg2019cellular,ratasuk2016nb}. Considering that most IoT devices do not require high transmission rates, LTE-M allocates a bandwidth of 1.4 MHz, while NB-IoT assigns 180 kHz to support cellular IoT~\cite{sharma2019toward}. Consequently, LTE-M provides mid-range coverage and supports higher transmission rates, making it suitable for voice and video services, whereas NB-IoT offers a broader coverage and is well-suited for low-rate devices.

For existing cellular IoT, numerous access schemes have been extensively researched to address the challenge of enabling massive IoT connections. In general, these multiple access schemes can be categorized into orthogonal and non-orthogonal approaches~\cite{shirvanimoghaddam2016multiple,shirvanimoghaddam2017massive,shirvanimoghaddam2017fundamental}. 
It was shown in~\cite{shirvanimoghaddam2016multiple} that the non-orthogonal multiple access (NOMA)-based scheme could achieve the highest achievable rate, while the frequency division multiple access (FDMA)-based scheme demonstrated comparable performance to the NOMA-based scheme in the coordinated scenario. 
Furthermore, an integrated massive beam-division multiple access (BDMA) framework for cellular IoT was investigated in~\cite{jia2019massive}. The IoT devices were divided into multiple clusters using a beamspace clustering method, and the BS utilized a common receive beam to recover the information from the IoT devices within the same beam while using successive interference cancellation (SIC) to decode the devices' information within the cluster. In~\cite{mao2021rate}, rate splitting multiple access (RSMA) for cellular IoT was studied, and two schemes, namely time partitioning and power partitioning, were proposed. The former served two groups of IoT devices over orthogonal time slots, while the latter served them within the same time slot in a non-orthogonal manner. Despite significant progress in supporting massive IoT connections with cellular IoT, it is worth pointing out that the devices in cellular IoT utilize active radio frequency units, resulting in relatively high circuit and transmission power consumption. Currently, most cellular IoT technologies operate at the power level of tens
to hundreds of milliwatts~\cite{garcia2023energy}. However, certain IoT devices are constrained by their physical size and can not accommodate the battery, leading to the extremely-low power consumption requirement in the order of microwatts~\cite{xu2021opportunities,3gppambient}. Existing cellular IoT techniques fail to support such devices and it is imperative to develop novel techniques that are more cost-effective to support massive IoT connections with ultra-low power consumption.




Recently, symbiotic radio (SR) has emerged as another efficient way to support massive IoT connections with cellular networks~\cite{liang2020symbiotic,liang2022symbiotic,janjua2021survey}. In conventional cellular IoT, the IoT devices utilize the cellular network for data transmission. In the SR system, there exist two types of transmissions, namely, the IoT and cellular transmissions. Specifically, the BS transmits information to the cellular user, while the IoT device realizes information transmission by backscattering the incident cellular signal. The receiver then jointly decodes the cellular and IoT information. Due to the passive reflecting nature, the IoT device in SR does not require power-hungry active components and is thus more cost-effective. Moreover, SR offers a two-fold advantage: On one hand, the cellular transmission provides an opportunity for IoT devices to transmit information passively without requiring additional power and spectrum resources; on the other hand, IoT transmission enhances cellular transmission by providing an additional multipath. The IoT and cellular transmissions thus form a mutualistic relationship, benefiting each other through their coexistence. As a result, SR is considered a promising technique to support massive access for cellular IoT in 6G networks~\cite{zhang20196g,huang2019survey}.

To facilitate massive IoT devices with SR, various multiple access schemes have been proposed, including code division multiple access (CDMA)~\cite{han2021design}, time division multiple access (TDMA)~\cite{yang2021energy,yeganeh2023multi}, and spatial division multiple access (SDMA)~\cite{chen2020stochastic} schemes. In the CDMA-based scheme presented in~\cite{han2021design}, each IoT device independently chose a random code to backscatter its information. 
In the TDMA-based scheme proposed in~\cite{yang2021energy}, IoT devices were assigned specific time slots to take turns transmitting information while harvesting energy from the cellular signal to support circuit operations. 
Additionally, a novel time-sharing method was proposed in~\cite{yeganeh2023multi}, where energy consumption was minimized under the IoT transmission rate requirement and energy harvesting constraints. The SDMA-based scheme in~\cite{chen2020stochastic} utilized multiple antennas at both BS and user sides, and the transmit and receive beamforming was designed to alleviate inter-IoT-device interference. Furthermore, the relationship between the achievable rates of cellular and IoT transmissions and the number of connected IoT devices was revealed in~\cite{wang2023multiple,xu2023mimo}. It was shown that as the number of connected IoT devices increases, higher transmission rates could be achieved. 
It is important to note that existing works primarily focus on single-user multi-IoT-device access schemes, and few have shed light on the scenario where multiple cellular users and multiple IoT devices coexist. As a matter of fact, the multi-user multi-IoT-device scenario is prevalent in reality. For example, in a smart home setting, multiple sensors may need to transmit sensing information (such as the temperature, humidity, and air quality) to multiple cellular users~\cite{kuai2021message}. Designing appropriate multiple access schemes for such practical scenarios is imminent. In addition, how to redesign the beamforming at the BS for such a cellular-IoT-integrated system is another challenge that needs to be addressed.


Motivated by the above, in this paper, we propose a novel multi-user multi-IoT-device SR system, in which the BS transmits information to multiple cellular users, and multiple IoT devices transmit information to the users by passively backscattering the incident cellular signal simultaneously. The information of different users and IoT devices are coupled with each other, and the received signals at the users are complicated due to the presence of inter-user and inter-IoT-device interference. 
Based on the proposed system, this paper aims to minimize the transmit power under the cellular transmission outage probability constraints and IoT transmission sum rate constraints by designing the beamforming vectors at BS. The main contributions of the paper are summarized as follows:
\begin{itemize}
    \item We propose a novel multi-user multi-IoT-device SR system, which serves as an innovative approach to support massive IoT connections with cellular networks. The multiple access design for such a cellular-IoT-integrated system is provided. The proposed system can achieve lower power consumption as compared to conventional cellular IoT systems.
    \item Noting that the reflective links from the IoT devices can be regarded as the channel uncertainty of the direct links, we apply the robust design method to design the beamforming vectors at BS. The transmit power is minimized under the cellular transmission outage probability constraints and IoT transmission sum rate constraints. The algorithm based on semi-definite programming (SDP), and difference-of-convex (DC) techniques is proposed to solve the formulated power minimization problem.
    \item Considering that the IoT devices are normally deployed around the cellular users, we investigate a special case of the multi-user multi-IoT-device SR system and propose a direction of arrival (DoA)-based transmit beamforming design approach. Specifically, due to the correlations between the reflective links and direct links, we use the DoAs and angular spreads (ASs) of the direct links instead of the instantaneous channel state information (CSI) of the reflective link channels to design the beamforming vectors and thus avoid the prohibitive channel feedback.
    \item The numerical results are provided to validate the multi-user multi-IoT-device SR system and the proposed CSI-based and DoA-based beamforming approaches. It is shown that the DoA-based beamforming approach achieves comparable performance as the CSI-based approach in the special case when the ASs are small.
\end{itemize}

The rest of the paper is organized as follows. In Section~\ref{system model}, we present the system model of multi-user multi-IoT-device SR systems and formulate the transmit power minimization problem. The optimization algorithm is developed in Section~\ref{sec_algorithm}. In Section~\ref{sec_specific}, we consider a special case and propose a DoA-based beamforming approach to reduce the channel feedback overhead in the considered system. Section~\ref{sec-simulation} presents the simulation results to verify the superiority of the proposed system and the effectiveness of the optimization algorithm. Finally, Section~\ref{sec-conclusion} concludes this paper.

The notations used in this paper are listed as follows. The lowercase, boldface lowercase, and boldface uppercase letters, e.g., $x$, $\boldsymbol{x}$, and $\boldsymbol{X}$ denote a scalar, vector, and matrix, respectively. $\mathcal{CN}(\mu,\sigma^2)$ denotes the circularly symmetric complex Gaussian (CSCG) distribution with mean $\mu$ and variance $\sigma^2$. $|x|$ denotes the operation of taking the absolute value of scalar $x$. $\text{Re}(x)$ denotes the real part of the complex scalar $x$. $\Vert\boldsymbol{x}\Vert$ denotes the operation of taking the $l_2$-norm value of the vector $\boldsymbol{x}$. $\boldsymbol{x}^H$ denotes the hermitian of vector $\boldsymbol{x}$. $\text{rank}(\boldsymbol{X})$ and $\text{tr}(\boldsymbol{X})$ denote the trace and rank of matrix $\boldsymbol{X}$. $\Vert \boldsymbol{X} \Vert_{*}$ and $\Vert \boldsymbol{X} \Vert_{2}$ denotes the nuclear norm and the spectral norm of matrix $\boldsymbol{X}$. $\boldsymbol{X}\succeq 0$ indicates that $\boldsymbol{X}$ is a positive semi-definite matrix. $\boldsymbol{I}_M$ denotes the $M$-dimensional identity matrix. $\mathbb{C}$ denotes the set of complex numbers and $\mathbb{C}^{M\times N}$ denotes the complex matrix with $M$ rows and $N$ columns.

\section{System Model}\label{system model}
\begin{figure}[!t]
\centering
\includegraphics[width=\columnwidth] {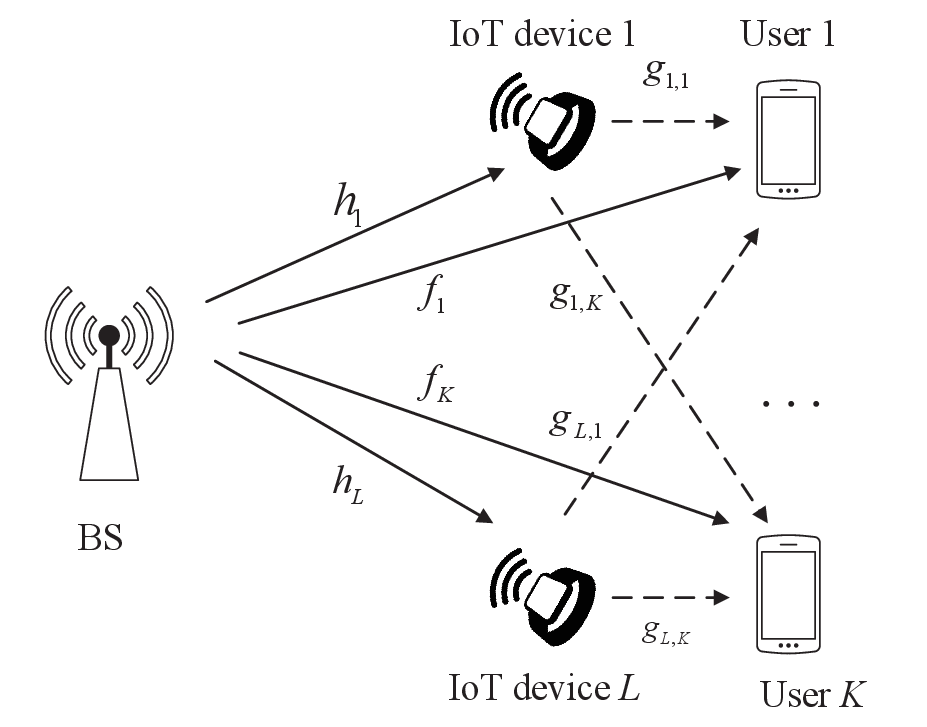}\vspace{-0.2cm}
\caption{System model for multi-user multi-IoT-device symbiotic radio systems.}
\label{fig:Fig1}
\vspace{-0.4cm}
\end{figure}

In this paper, we consider a multi-user multi-IoT-device symbiotic radio system. As illustrated in Fig.~\ref{fig:Fig1}, the considered system consists of a BS with $M$ antennas, $K$ single-antenna cellular users, and $L$ single-antenna IoT devices. The channels from the BS to the $k$-th user, from the BS to the $l$-th IoT device, and from the $l$-th IoT device to the $k$-th user are denoted by $\boldsymbol{f}_k\in \mathbb{C}^{1\times M}$, $\boldsymbol{h}_l\in \mathbb{C}^{1\times M}$, and $\boldsymbol{g}_{l,k}\in \mathbb{C}$, respectively. In this section, we introduce the signal model of the proposed system, discuss the channel estimation and synchronization approach, and formulate the transmit power minimization problem.
\subsection{Signal Model}
The BS transmits information to multiple cellular users and the transmit signal is given by
\begin{align}
    \boldsymbol{s}(n)=\sum\nolimits_{i=1}^K\boldsymbol{w}_is_i(n),
\end{align}
where $s_i(n)\sim \mathcal{CN}(0,1)$ and $\boldsymbol{w}_i\in \mathbb{C}^{M \times 1}$ denote the intended symbol and the corresponding beamforming vector for cellular user $i$. Multiple IoT devices transmit their information to cellular users simultaneously by passively backscattering the incident cellular signal. The signal received by user $k$ consisting of two parts, the direct link signal from the BS and the reflective link signals from the IoT devices, is written as
\begin{align}
{y_k}(n) = \boldsymbol{f}_k\boldsymbol{s}(n) + \sum\nolimits_{l = 1}^L {\alpha {g_{l,k}}{c_l}{\boldsymbol{h}_l}\boldsymbol{s}(n)}  + {u_k}(n),
\end{align}
where $c_l\sim \mathcal{CN}(0,1)$ denotes the transmit symbol of IoT device $l$, $\alpha$ denotes the reflection coefficient of IoT devices, and ${u_k}(n)$ is the additive white Gaussian noise (AWGN) following the distribution of $\mathcal{CN}(0,\sigma_k^2)$. The symbol period of IoT transmission symbol $c_l$ is $N$ times that of the BS symbol $\boldsymbol{s}(n)$. User $k$ adopts SIC to decode its information $s_k(n)$ from BS and the information $\left\{c_l\right\}_{l=1}^L$ from IoT devices. Since the backscatter link is relatively weak compared with the direct link due to the double fading effect, user $k$ first decodes its information $s_k(n)$. The received signal can be rewritten as
\begin{align}
    {y_k}(n) &\!=\! \left( {{\boldsymbol{f}_k} \!+\! \sum\nolimits_{l = 1}^L {\alpha {g_{l,k}}{c_l}\boldsymbol{h}_l} } \right){\boldsymbol{w}_k}{s_k}(n)  \nonumber\\
    &\!+\!\left( {{\boldsymbol{f}_k} \!+\! \sum\nolimits_{l = 1}^L {\alpha {g_{l,k}}{c_l}\boldsymbol{h}_l} } \right)\sum\nolimits_{i = 1,i \ne k}^K {{\boldsymbol{w}_i}{s_i}(n)}  \!+\! {u_k}(n).\label{y_k}
\end{align}
When decoding $s_k(n)$, the first item in (\ref{y_k}) is the desired signal, while the second item is the interference signal. The signal-to-interference-plus-noise ratio (SINR) for $s_k(n)$ is thus given by
\begin{align}
{\gamma_{s,k}} = \frac{{{{\left| {\left( {{\boldsymbol{f}_k} + \sum\nolimits_{l = 1}^L {\alpha {g_{l,k}}{c_l}{\boldsymbol{h}_l}} } \right){\boldsymbol{w}_k}} \right|}^2}}}{{\sum\nolimits_{i = 1,i \ne k}^K {{{\left| {\left( {{\boldsymbol{f}_k} + \sum\nolimits_{l = 1}^L {\alpha {g_{l,k}}{c_l}{\boldsymbol{h}_l}} } \right){\boldsymbol{w}_i}} \right|}^2}}  + \sigma _k^2}}.
\end{align}
The instantaneous rate for $s_k(n)$ at user $k$ is given by
\begin{align}
{R_{s,k}} = {{{\log }_2}\left( {1 + {\gamma _{s,k}}} \right)}.
\end{align}
Note that the instantaneous rate contains the information $\left\{c_l\right\}_{l=1}^L$ from IoT devices, which changes relatively fast as compared to the channel variation. The actual value of $c_l$ is not known to the receiver when decoding $s_k(n)$, but the knowledge of the distribution of $c_l$ is available. Inspired by the robust beamforming approach investigated in~\cite{medra2016low,zheng2010robust}, in this paper, we treat the reflective links from the IoT devices containing $\left\{c_l\right\}_{l=1}^L$ as the channel uncertainty of the direct link, and utilize the statistical approach to deal with it. The reasons for adopting this approach are as follows: Firstly, the reflective links exhibit significantly lower strength than the direct link due to the double fading effect, which is similar to the characteristic of the channel estimation error. Secondly, the distribution of the reflective link can be derived with the distribution of $c_l$ known, and thus the statistical approach can be applied.
We denote the reflective link as $\Delta\boldsymbol{f}_k=\sum\nolimits_{l = 1}^L {\alpha {g_{l,k}}{c_l}{\boldsymbol{h}_l}}\in \mathbb{C}^{1 \times M}$. Since $c_l$ follows the distribution of $\mathcal{CN}(0,1)$, and $g_{l,k}$ and $\boldsymbol{h}_l$ are perfectly known, the reflective link $\Delta\boldsymbol{f}_k$ follows the distribution of $\mathcal{CN}(\boldsymbol{0},\boldsymbol{C}_k)$, where the covariance matrix $\boldsymbol{C}_k$ is calculated as
\begin{align}
    \boldsymbol{C}_k=\sum\nolimits_{l = 1}^L {|\alpha {g_{l,k}}{|^2}\boldsymbol{h}_l^H{\boldsymbol{h}_l}}.
\end{align}
Considering the effect of $\left\{c_l\right\}_{l=1}^L$, we define the outage event for cellular transmission as follows: the outage event occurs when the instantaneous cellular transmission rate is lower than the rate requirement $\bar{{R}}_{s,k}$ and the outage probability of the cellular transmission for user $k$ is thus given by
\begin{align}
    P_{\text{out},k}=\text{Pr}\left\{{R_{s,k}}\leq \bar{{R}}_{s,k}\right\}.
\end{align}
We should note that the occurrence of the outage event is related to the direct link channels, as well as the channel uncertainties, i.e., the reflective link channels. 
\begin{figure}[!t]
\centering
\includegraphics[width=\columnwidth] {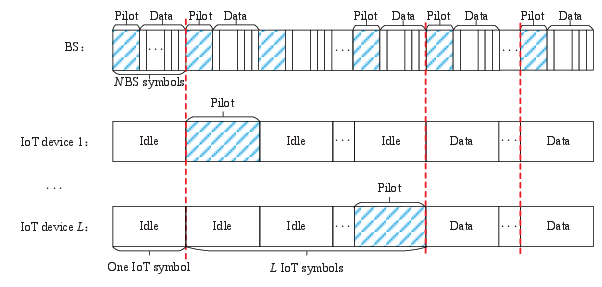}
\caption{An illustration of the timeline for multi-user multi-IoT-device SR systems.}
\label{fig:Fig_frame}
\end{figure}

After decoding its information $s_k(n)$, user $k$ subtracts the known direct link signal  $\boldsymbol{f}_k\boldsymbol{w}_ks_k(n)$ and decodes the information of IoT devices by using the SIC technique. The received signal after direct link cancellation is given by
\begin{align}
    {\bar{y}_k}(n) &=   \sum\nolimits_{l = 1}^L {\alpha {g_{l,k}}{c_l}\boldsymbol{h}_l}  {\boldsymbol{w}_k}{s_k}(n)  \nonumber\\
    &\!+\!\left( {{\boldsymbol{f}_k} \!+\! \sum\nolimits_{l = 1}^L {\alpha {g_{l,k}}{c_l}\boldsymbol{h}_l} } \right)\sum\nolimits_{i = 1,i \ne k}^K {{\boldsymbol{w}_i}{s_i}(n)}  \!+\! {u_k}(n).
\end{align}
When user $k$ decodes the information $c_m$ from IoT device $m$, the interference caused by the first $m-1$ devices, i.e., $\left\{c_l\right\}_{l=1}^{m-1}$ can be cancelled, and the signal when decoding $c_m$ can be expressed by
\begin{align}
    {{y}_{m,k}}(n) &\!=\!   \alpha {g_{m,k}}{c_m}\boldsymbol{h}_m  {\boldsymbol{w}_k}{s_k}(n)\!+\!\sum\nolimits_{l = m+1}^L {\alpha {g_{l,k}}{c_l}\boldsymbol{h}_l}  {\boldsymbol{w}_k}{s_k}(n) \nonumber\\
    &\!+\! \sum\nolimits_{l = 1}^L  \sum\nolimits_{i = 1,i \ne k}^K{\alpha {g_{l,k}}{c_l}\boldsymbol{h}_l} {{\boldsymbol{w}_i}{s_i}(n)}\nonumber\\
    &\!+\!{\boldsymbol{f}_k}\sum\nolimits_{i = 1,i \ne k}^K {{\boldsymbol{w}_i}{s_i}(n)}
      \!+\! {u_k}(n).\label{y_tilde}
\end{align}
In (\ref{y_tilde}), the first item is the desired signal, the second item is the interference signal from the remaining $L-m$ devices with the same $s_k(n)$, the third item is the interference signal from other devices with different $s_k(n)$, and the fourth item is the residual direct link interference. Since the symbol period of the IoT devices covers $N$ BS symbol periods, the maximum-ratio combining (MRC) technique can be utilized to combine $N$ symbols of $y_k(n)$. Therefore, the SINR for user $k$ to decode the information $c_m$ with MRC is written as (\ref{gamma_c_m}), shown at the bottom of the next page.
\begin{figure*}[b]
    \begin{center}
    \begin{align}
{\gamma _{{c_m},k}} = \frac{{N{{\left| {\alpha {g_{m,k}}{\boldsymbol{h}_m}{\boldsymbol{w}_k}} \right|}^2}}}{{\sum\nolimits_{i = 1,i \ne k}^K {{{\left| {{\boldsymbol{f}_k}{\boldsymbol{w}_i}} \right|}^2}}  + \sum\nolimits_{l = 1}^L {\sum\nolimits_{i = 1,i \ne k}^K {{{\left| {\alpha {g_{l,k}}{\boldsymbol{h}_l}{\boldsymbol{w}_i}} \right|}^2}} }  + N\sum\nolimits_{l{\rm{ = }}m{\rm{ + }}1}^L {{{\left| {\alpha {g_{l,k}}{\boldsymbol{h}_l}{\boldsymbol{w}_k}} \right|}^2}}  + \sigma _k^2}}.\label{gamma_c_m}
\end{align}
\end{center}
\end{figure*}

The corresponding achievable rate can be expressed as
\begin{align}
{R_{{c_m},k}} = \frac{1}{N}{\log_2}\left( {1 + {\gamma _{{c_m},k}}} \right).
\end{align}
The sum IoT device rate for user $k$ is accordingly given by (\ref{R_k_c}), shown at the bottom of the next page.
\begin{figure*}[b]
    \begin{center}
    \begin{align}
{R _{c,k}} = \frac{1}{N}{\log_2}\left( {1 +\frac{N\sum\nolimits_{l = 1}^L {{{\left| {\alpha {g_{l,k}}{\boldsymbol{h}_l}{\boldsymbol{w}_k}} \right|}^2}}}{{\sum\nolimits_{i = 1,i \ne k}^K {{{\left| {{\boldsymbol{f}_k}{\boldsymbol{w}_i}} \right|}^2}}  + \sum\nolimits_{l = 1}^L {\sum\nolimits_{i = 1,i \ne k}^K {{{\left| {\alpha {g_{l,k}}{\boldsymbol{h}_l}{\boldsymbol{w}_i}} \right|}^2}} }  + \sigma _k^2}}}\right).\label{R_k_c}
\end{align}
\end{center}
\end{figure*}


\subsection{Channel Estimation and Synchronization Discussions}

In this paper, we have assumed that the perfect CSI is available at the BS and users via channel estimation techniques. An illustration of the timeline is depicted in Fig.~\ref{fig:Fig_frame}. As illustrated in Fig.~\ref{fig:Fig_frame}, the channel estimation process can be divided into two phases. In both phases, the BS transmits pilot signals, and the users perform channel estimation. In the first phase (the first IoT symbol), the IoT devices stay idle by tuning the impedance into the matched state so that the users can estimate the direct link channel $\boldsymbol{f}_k$ based on the received pilot signals~\cite{liang2022backscatter}. In the second phase (the following $L$ IoT symbols), each IoT device transmits the pilot symbol in turns by tuning the impedance into the backscattering state while the others remain idle at the same time. The users receive the direct link signal from the BS and the reflective link signal from the specific IoT device, and the reflective channels $g_{l,k}\boldsymbol{h}_l,\forall l$ can be estimated based on the received signals by subtracting the direct link channel from the estimated composite channel $\boldsymbol{f}_k+g_{l,k}\boldsymbol{h}_l$. With the obtained CSI, data transmission and signal detection can then be performed. During the data transmission procedure, the BS also transmits pilot signals to remove the ambiguity between $s(n)$ and $c_l$ in case of the direct link is weak or blocked~\cite{zhang2021reconfigurable}.

Moreover, similar to the earlier works on SR~\cite{chen2020stochastic,zhang2021reconfigurable}, we have assumed perfect synchronization between the signal of the BS and those of the IoT devices, as shown in Fig.~\ref{fig:Fig_frame}. The reasons are as follows. Firstly, since the IoT device only reflects its incident signal passively and there is no signal processing or receiving procedure along the reflection paths, the delay caused by the IoT device's backscattering is negligible. Secondly, due to the limited transmission range, the distances between IoT devices and users are assumed to be much shorter than those between BS and users, leading to the same transmission delay between direct and reflective links. Furthermore, several time synchronization techniques have been studied to accurately estimate and compensate for the time delay, such as~\cite{zhou2019cortis,morelli2007synchronization}.


\subsection{Problem Formulation}
To further investigate the performance of the proposed multi-user multi-IoT-device SR system, in this paper, we minimize the transmit power at the BS under the cellular transmission outage probability constraints and IoT transmission sum rate constraints by designing the transmit beamforming vectors at the BS. The transmit power minimization problem is formulated as
\begin{subequations}
\begin{align}
{\bf (P1) }:\underset{\left\{\boldsymbol{w}_k\right\}_{k=1}^K}{\min}&\quad \sum\nolimits_{k=1}^K\Vert \boldsymbol{w}_k\Vert^2\nonumber \\
\text{s.t.}\quad
&\text{Pr}\left\{{R_{s,k}}\leq \bar{{R}}_{s,k}\right\}\leq P_{\text{out}},\forall k,\label{cellular outage cons}\\
&R _{c,k}\geq \bar{R}_{c,k},\forall k,\label{IoT cons}
\end{align}
\end{subequations}
where $P_{\text{out}}$ is the predetermined cellular transmission outage requirement, and $\bar{R}_{c,k}$ is the sum rate requirement for IoT transmission at user $k$. Constraint (\ref{cellular outage cons}) indicates that the outage probability of cellular transmission for each user should be less than the threshold $P_{\text{out}}$, which guarantees the quality of service of cellular transmission. Constraint (\ref{IoT cons}) represents that the sum rate of IoT transmission should be higher than the predetermined value $\bar{R}_{c,k}$, ensuring the quality of service of IoT transmission.

\section{Proposed Algorithm}\label{sec_algorithm}
In this section, we propose an optimization algorithm to solve the transmit power minimization problem formulated in previous section.
\subsection{Problem Transformation}
It is challenging to solve problem ${\bf (P1)}$ due to the non-convex outage constraint (\ref{cellular outage cons}). In the following, we will transform the constraint into a tractable one using $\mathcal{S}$-lemma. The outage constraint (\ref{cellular outage cons}) can be explicitly written as
\begin{align}
    \text{Pr}&\left\{{{{\log }_2}\left( {1 + \frac{{{{\left| {\left( {{\boldsymbol{f}_k} + \Delta\boldsymbol{f}_k } \right){\boldsymbol{w}_k}} \right|}^2}}}{{\sum\nolimits_{i = 1,i \ne k}^K {{{\left| {\left( {{\boldsymbol{f}_k} + \Delta\boldsymbol{f}_k } \right){\boldsymbol{w}_i}} \right|}^2}}  + \sigma _k^2}}} \right)} \!\leq\! \bar{{R}}_{s,k}\right\} \nonumber\\
    &\qquad\qquad\qquad\qquad\qquad\qquad\qquad\qquad\qquad\leq P_{\text{out}}.\label{out_left}
\end{align}
We can rewrite (\ref{out_left}) as
\begin{align}
\text{Pr} \Bigg\{ &
{\Delta {\boldsymbol{f}_k}\left( {\frac{{{\boldsymbol{w}_k}\boldsymbol{w}_k^H}}{{{2^{{{\bar R }_{s,k}}}} - 1}} - \sum\nolimits_{i = 1,i \ne k}^K {{\boldsymbol{w}_i}\boldsymbol{w}_i^H} } \right)\Delta \boldsymbol{f}_k^H}+\nonumber\\
&{2{\mathop{\rm Re}\nolimits} \left[ {\Delta {\boldsymbol{f}_k}{{\left( {\frac{{{\boldsymbol{w}_k}\boldsymbol{w}_k^H}}{{{2^{{{\bar R }_{s,k}}}} - 1}} - \sum\nolimits_{i = 1,i \ne k}^K {{\boldsymbol{w}_i}\boldsymbol{w}_i^H} } \right)}^H}\boldsymbol{f}_k^H} \right]}+\nonumber\\
&{{\boldsymbol{f}_k}\left( {\frac{{{\boldsymbol{w}_k}\boldsymbol{w}_k^H}}{{{2^{{{\bar R }_{s,k}}}} - 1}} - \sum\nolimits_{i = 1,i \ne k}^K {{\boldsymbol{w}_i}\boldsymbol{w}_i^H} } \right)\boldsymbol{f}_k^H - \sigma _k^2 \le 0}\Bigg\}\nonumber\\
&\leq P_{\text{out}}.\label{Pr}
\end{align}
We can further simplify (\ref{Pr}) as
\begin{align}
\Pr \left\{ {\boldsymbol{e}{\boldsymbol{Q}_k}{\boldsymbol{e}^H} + 2{\mathop{\rm Re}\nolimits} \left\{ {\boldsymbol{e}{\boldsymbol{r}_k}} \right\} + {s_k} \le 0} \right\}\leq P_{\text{out}}.\label{cons_sim}
\end{align}
where 
\begin{align}
{\boldsymbol{Q}_k} &= \boldsymbol{C}_k^{\frac{1}{2}}\left( {\frac{{{\boldsymbol{w}_k}\boldsymbol{w}_k^H}}{{{2^{{{\bar R }_{s,k}}}} - 1}} - \sum\nolimits_{i = 1,i \ne k}^K {{\boldsymbol{w}_i}\boldsymbol{w}_i^H} } \right)\boldsymbol{C}_k^{\frac{1}{2}},\label{Q_k_value}\\ 
{\boldsymbol{r}_k} &= \boldsymbol{C}_k^{\frac{1}{2}}{\left( {\frac{{{\boldsymbol{w}_k}\boldsymbol{w}_k^H}}{{{2^{{{\bar R }_{s,k}}}} - 1}} - \sum\nolimits_{i = 1,i \ne k}^K {{\boldsymbol{w}_i}\boldsymbol{w}_i^H} } \right)^H}\boldsymbol{f}_k^H,\label{r_k_value} \\ 
{s_k} &= {\boldsymbol{f}_k}\left( {\frac{{{\boldsymbol{w}_k}\boldsymbol{w}_k^H}}{{{2^{{{\bar R }_{s,k}}}} - 1}} - \sum\nolimits_{i = 1,i \ne k}^K {{\boldsymbol{w}_i}\boldsymbol{w}_i^H} } \right)\boldsymbol{f}_k^H - \sigma _k^2,\label{s_k_value}
\end{align}
and $\boldsymbol{e}\in \mathbb{C}^{1 \times M}$ is a standard complex Gaussian vector (i.e., $\boldsymbol{e}\sim \mathcal{CN}(\boldsymbol{0},\boldsymbol{I}_M)$). The constraint (\ref{cons_sim}) is equivalent to
\begin{align}
\Pr \left\{ {\boldsymbol{e}{\boldsymbol{Q}_k}{\boldsymbol{e}^H} + 2{\mathop{\rm Re}\nolimits} \left\{ {\boldsymbol{e}{\boldsymbol{r}_k}} \right\} + {s_k} \ge 0} \right\}\geq 1-P_{\text{out}}.\label{cons_sim_greater}
\end{align}
To deal with the outage probability constraint, we have the following proposition:
\begin{proposition}
Suppose there is a set $\mathcal{A}$ that satisfies $\text{Pr}\left\{\boldsymbol{e}\in A\right\}\geq 1-\rho$, and the condition ${\boldsymbol{\eta}{\boldsymbol{Q}}{\boldsymbol{\eta}^H} + 2{\mathop{\rm Re}\nolimits} \left\{ {\boldsymbol{\eta}{\boldsymbol{r}}} \right\} + {s} \ge 0},\forall \boldsymbol{\eta} \in \mathcal{A}$ holds, we can conclude that the constraint (\ref{cons_sim_greater}) holds.
\begin{IEEEproof}
The proof can be found in~\cite{wang2014outage}.
\end{IEEEproof}
\end{proposition}
The set $\mathcal{A}$ can be chosen to be a spherical set, i.e., 
\begin{align}
    \mathcal{A}=\left\{\boldsymbol{\eta} \in \mathbb{C}^M| \Vert\boldsymbol{\eta} \Vert \leq d\right\},
\end{align}
where $d=\sqrt {\frac{{\Phi _{\chi _{2M}^2}^{ - 1}\left( {1 - \rho} \right)}}{2}}$ denotes the sphere radius, and $\Phi _{\chi _{q}^2}^{ - 1}(\cdot)$ is the inverse cumulative distribution function (CDF) of the central Chi-square random variable with $q$ degrees of freedom. By using the $\mathcal{S}$-lemma, we can transform the probabilistic constraint into a tractable constraint. The $\mathcal{S}$-lemma is given as follows.
\begin{lemma}
Let $f(\boldsymbol{x})=\boldsymbol{x}{\boldsymbol{Q}_1}\boldsymbol{x}^H+2{\mathop{\rm Re}\nolimits} \left\{ {\boldsymbol{x}{\boldsymbol{r}_1}} \right\} + {s}_1$ and $g(\boldsymbol{x})=\boldsymbol{x}{\boldsymbol{Q}_2}\boldsymbol{x}^H+2{\mathop{\rm Re}\nolimits} \left\{ {\boldsymbol{x}{\boldsymbol{r}_2}} \right\} + {s}_2$, where $\boldsymbol{Q}_1$ and $\boldsymbol{Q}_2$ are Hermitian matrices, $\boldsymbol{r}_1$ and $\boldsymbol{r}_2$ are vectors, and $s_1$ and $s_2$ are real numbers. Assume that there is an $\boldsymbol{x}_0 \in \mathbb{C}^n$ such that $g(\boldsymbol{x}_0)<0$ holds. The following two statements are always equivalent:
\begin{itemize}
    \item $\forall \boldsymbol{x} \in \mathbb{C}^n$, $g(\boldsymbol{x})\leq 0 \Rightarrow f(\boldsymbol{x}) \geq 0$.
    \item There exists a $\mu\geq 0$ such that 
    \begin{align}
 \begin{bmatrix}
   \boldsymbol{Q}_1 & \boldsymbol{r}_1  \\
   \boldsymbol{r}_1^H & {s}_1 
  \end{bmatrix}+ \mu 
  \begin{bmatrix}
   \boldsymbol{Q}_2 & \boldsymbol{r}_2  \\
   \boldsymbol{r}_2^H & {s}_2 
  \end{bmatrix}\succeq 0.
\end{align}
\end{itemize}
\end{lemma}
With $f(\boldsymbol{x})=\boldsymbol{x}{\boldsymbol{Q}_k}\boldsymbol{x}^H+2{\mathop{\rm Re}\nolimits} \left\{ {\boldsymbol{x}{\boldsymbol{r}_k}} \right\} + {s}_k$ and $g(\boldsymbol{x})=\boldsymbol{x}\boldsymbol{x}^H-d^2$, by using $\mathcal{S}$-lemma, we can transform the outage constraint (\ref{cons_sim_greater}) to
\begin{align}
    \begin{bmatrix}
   \boldsymbol{Q}_k+\mu_k \boldsymbol{I}_M & \boldsymbol{r}_k  \\
   \boldsymbol{r}_k^H & {s}_k-\mu_k d^2
  \end{bmatrix}\succeq 0,
\end{align}
where $\mu_k$ is the slack variable. The problem ${\bf (P1)}$ is thus rewritten as
\begin{subequations}
\begin{align}
{\bf (P2) }:\underset{\left\{\boldsymbol{w}_k,\mu_k\right\}_{k=1}^K}{\min}&\quad \sum\nolimits_{k=1}^K\Vert \boldsymbol{w}_k\Vert^2\nonumber \\
\text{s.t.}\quad
&\begin{bmatrix}
   \boldsymbol{Q}_k+\mu_k \boldsymbol{I}_M & \boldsymbol{r}_k  \\
   \boldsymbol{r}_k^H & {s}_k-\mu_k d^2
  \end{bmatrix}\succeq 0,\forall k,\\ 
  &\mu_k\geq 0,\forall k,\label{mu_k}\\
&(\text{\ref{IoT cons}}), (\text{\ref{Q_k_value}}), (\text{\ref{r_k_value}}), \text{and} \ (\text{\ref{s_k_value}}).\nonumber
\end{align}
\end{subequations}
Furthermore, the constraint $(\text{\ref{IoT cons}})$ is explicitly expressed as
\begin{align}
    &\frac{N\sum\nolimits_{l = 1}^L {{{\left| {\alpha {g_{l,k}}{\boldsymbol{h}_l}{\boldsymbol{w}_k}} \right|}^2}}}{{\sum_{i = 1,i \ne k}^K {{{\left| {{\boldsymbol{f}_k}{\boldsymbol{w}_i}} \right|}^2}}  \!+\! \sum_{l = 1}^L {\sum_{i = 1,i \ne k}^K {{{\left| {\alpha {g_{l,k}}{\boldsymbol{h}_l}{\boldsymbol{w}_i}} \right|}^2}} }  \!+\! \sigma _k^2}} \nonumber\\
    &\qquad\qquad\qquad\qquad\qquad\qquad\qquad\qquad\geq 2^{N\bar{R}_{c,k} }-1.
\end{align}
Denote $\boldsymbol{\psi}_{l,k}=\alpha {g_{l,k}}{\boldsymbol{h}_l}$ and $\bar{\gamma}_{c,k}=2^{N\bar{R}_{c,k} }-1$, the constraint can be rewritten as
\begin{align}
    &\frac{N\sum\nolimits_{l = 1}^L {{{ {\boldsymbol{\psi}_{l,k}{\boldsymbol{w}_k}{\boldsymbol{w}_k^H}\boldsymbol{\psi}_{l,k}^H} }}}}{{\sum\nolimits_{i = 1,i \ne k}^K {{{ {{\boldsymbol{f}_k}{\boldsymbol{w}_i}\boldsymbol{w}_i^H\boldsymbol{f}_k^H} }}}  \!+ \!\sum\nolimits_{l = 1}^L {\sum\nolimits_{i = 1,i \ne k}^K {{{ {\boldsymbol{\psi}_{l,k}{\boldsymbol{w}_i}\boldsymbol{w}_i^H\boldsymbol{\psi}_{l,k}^H} }}} } \! +\! \sigma _k^2}} \nonumber\\
    &\qquad\qquad\qquad\qquad\qquad\qquad\qquad\qquad\qquad\geq \bar{\gamma}_{c,k}.\label{iot_cons}
\end{align}
We can transform the constraint (\ref{iot_cons}) to
\begin{align}
    &\sum\nolimits_{l = 1}^L {{{ {\boldsymbol{\psi}_{l,k}\left(\frac{N\boldsymbol{w}_k\boldsymbol{w}_k^H}{\bar{\gamma}_{c,k}}-\sum\nolimits_{i = 1,i \ne k}^K{\boldsymbol{w}_i}\boldsymbol{w}_i^H\right)\boldsymbol{\psi}_{l,k}^H} }}}\nonumber\\
    &-\sum\nolimits_{i = 1,i \ne k}^K {{{ {{\boldsymbol{f}_k}{\boldsymbol{w}_i}\boldsymbol{w}_i^H\boldsymbol{f}_k^H} }}}- \sigma _k^2\geq 0.\label{cons2}
\end{align}
Noting $\boldsymbol{C}_k=\sum\nolimits_{l = 1}^L\boldsymbol{\psi}_{l,k}^H\boldsymbol{\psi}_{l,k}$, the constraint (\ref{cons2}) can be further transformed to
\begin{align}
    &\text{tr}\left( {{ {\left(\frac{N\boldsymbol{w}_k\boldsymbol{w}_k^H}{\bar{\gamma}_{c,k}}-\sum\nolimits_{i = 1,i \ne k}^K{\boldsymbol{w}_i}\boldsymbol{w}_i^H\right)\boldsymbol{C}_k }}}\right)\nonumber\\
    &-\sum\nolimits_{i = 1,i \ne k}^K {{{ {{\boldsymbol{f}_k}{\boldsymbol{w}_i}\boldsymbol{w}_i^H\boldsymbol{f}_k^H} }}}- \sigma _k^2\geq 0.\label{cons2}
\end{align}

Denote $\boldsymbol{W}_k=\boldsymbol{w}_k\boldsymbol{w}_k^H$, which is equivalent to $\text{rank}\left(\boldsymbol{W}_k\right)=1,\boldsymbol{W}_k \succeq 0$. The problem ${\bf (P2)}$ can be transformed to the following problem ${\bf (P3)}$ with respect to $\boldsymbol{W}_k$:
\begin{subequations}
\begin{align}
{\bf (P3) }&:\underset{\left\{\boldsymbol{W}_k,\mu_k\right\}_{k=1}^{K}}{\min}\quad \sum\nolimits_{k=1}^K\text{tr}(\boldsymbol{W}_k)\nonumber \\
\text{s.t.}\quad
&\begin{bmatrix}
   \bar{\boldsymbol{Q}}_k+\mu_k \boldsymbol{I}_M & \bar{\boldsymbol{r}}_k  \\
   \bar{\boldsymbol{r}}_k^H & \bar{{s}}_k-\mu_k d^2
  \end{bmatrix}\succeq 0,\forall k,\label{p3-1}\\
&\text{tr}\left( {{ {\left(\frac{N\boldsymbol{W}_k}{\bar{\gamma}_{c,k}}-\sum\nolimits_{i = 1,i \ne k}^K{\boldsymbol{W}_i}\right)\boldsymbol{C}_k }}}\right)\nonumber\\
&-\sum\nolimits_{i = 1,i \ne k}^K { \text{tr}\left(\boldsymbol{W}_i\boldsymbol{f}_k^H\boldsymbol{f}_k \right)}- \sigma _k^2\geq 0,\forall k,\label{cons2_trans}\\
&\text{rank}\left(\boldsymbol{W}_k\right)=1, \forall k,\label{rank_cons}\\
&\boldsymbol{W}_k \succeq 0,\forall k,\label{semide_cons}\\
&\text{(\ref{mu_k})},\nonumber
\end{align}
\end{subequations}
where 
\begin{align}
{\bar{\boldsymbol{Q}}_k} &= \boldsymbol{C}_k^{\frac{1}{2}}\left( {\frac{\boldsymbol{W}_k}{{{2^{{{\bar R }_{s,k}}}} - 1}} - \sum\nolimits_{i = 1,i \ne k}^K \boldsymbol{W}_i } \right)\boldsymbol{C}_k^{\frac{1}{2}},\label{Q_k_bar}\\ 
{\bar{\boldsymbol{r}}_k} &= \boldsymbol{C}_k^{\frac{1}{2}}{\left( {\frac{\boldsymbol{W}_k}{{{2^{{{\bar R }_{s,k}}}} - 1}} -  \sum\nolimits_{i = 1,i \ne k}^K \boldsymbol{W}_i } \right)^H}\boldsymbol{f}_k^H,\label{r_k_bar} \\ 
\bar{{s}}_k &= {\boldsymbol{f}_k}\left( {\frac{\boldsymbol{W}_k}{{{2^{{{\bar R }_{s,k}}}} - 1}} - \sum\nolimits_{i = 1,i \ne k}^K \boldsymbol{W}_i } \right)\boldsymbol{f}_k^H - \sigma _k^2.\label{s_k_bar}
\end{align}
Note that the rank constraint (\ref{rank_cons}) is hard to tackle. The problem with a rank constraint is normally solved with the semi-definite relaxation (SDR) technique. In particular, the rank constraint in problem ${\bf (P3)}$ is first dropped, and the rank-relaxed problem, which is an SDP problem, is solved with CVX~\cite{grant2014cvx}. The obtained optimal solution is denoted by $\boldsymbol{W}_k^{\star}$. If the rank of $\boldsymbol{W}_k^{\star}$ is one, we perform eigenvalue decomposition for $\boldsymbol{W}_k^{\star}$ as $\boldsymbol{W}_k^{\star}=\boldsymbol{U}_k^H\Sigma_k\boldsymbol{U}_k$. The optimal beamforming vector can be given by $\boldsymbol{w}_k^{\star}=\boldsymbol{U}_k^H\boldsymbol{\Sigma}_k^{\frac{1}{2}}$. If the rank of $\boldsymbol{W}_k^{\star}$ is not one, the Gaussian randomization procedure is applied to find a rank-one solution~\cite{luo2010semidefinite}. Specifically, we generate a set of candidate random vectors $\boldsymbol{w}_k=\boldsymbol{U}_k^H\boldsymbol{\Sigma}_k^{\frac{1}{2}}\boldsymbol{e}_k$, where $\boldsymbol{e}_k \sim \mathcal{CN}(\boldsymbol{0},\boldsymbol{I}_M)$, and test whether the constraints can be satisfied by substituting the candidate $\boldsymbol{w}_k$. The candidate random vector that minimizes the objective function value is selected as the optimal solution. 

However, dropping the rank constraints will incur performance degradation, and the SDR technique with the Gaussian randomization procedure may fail to return a feasible solution. A unified difference-of-convex (DC) programming method was proposed in~\cite{fu2021reconfigurable} to address this problem, and it was shown that the DC method shows faster convergence behavior and better performance than the SDR method. Therefore, in this paper, we apply the DC programming method to find the rank-one solution. Since the matrices $\left\{\boldsymbol{W}_k\right\}_{k=1}^K$ are positive semi-definite matrices and $\text{tr}\left(\boldsymbol{W}_k\right)>0,\forall k$, the rank constraint (\ref{rank_cons}) can be rewritten as the following DC representation~\cite{fu2021reconfigurable}:
\begin{align}
    \Vert \boldsymbol{W}_k \Vert_{*}-\Vert \boldsymbol{W}_k \Vert_{2}=0,\forall k,
\end{align}
where $\Vert \boldsymbol{W}_k \Vert_{*}$ and $\Vert \boldsymbol{W}_k \Vert_{2}$ denotes the nuclear norm and the spectral norm of matrix $\boldsymbol{W}_k$. By introducing the penalty item to replace the rank constraint, the problem ${\bf (P3)}$ can be recast as
\begin{subequations}
\begin{align}
&{\bf (P4) }:\nonumber\\
&\underset{\left\{\boldsymbol{W}_k,\mu_k\right\}_{k=1}^{K}}{\min}\quad \sum\nolimits_{k=1}^K\text{tr}(\boldsymbol{W}_k)\!+\!\rho\sum\nolimits_{k=1}^K\left(\Vert \boldsymbol{W}_k \Vert_{*}\!-\!\Vert \boldsymbol{W}_k \Vert_{2}\right)\nonumber \\
&\quad\quad\quad\text{s.t.}\quad\
\text{(\ref{mu_k})},\text{(\ref{p3-1})},\text{(\ref{cons2_trans})},\ \text{and} \ \text{(\ref{semide_cons})},\nonumber
\end{align}
\end{subequations}
where $\rho>0$ is the penalty ratio. We can further rewrite the objective function in problem ${\bf (P4)}$ as the difference of two strongly convex functions as follows:
\begin{align}
    &\sum\nolimits_{k=1}^K\text{tr}(\boldsymbol{W}_k)\!+\!\rho\sum\nolimits_{k=1}^K\left(\Vert \boldsymbol{W}_k \Vert_{*}\!-\!\Vert \boldsymbol{W}_k \Vert_{2}\right)\nonumber \\
    =&\underbrace{\sum\nolimits_{k=1}^K\text{tr}(\boldsymbol{W}_k)+\rho\sum\nolimits_{k=1}^K\Vert \boldsymbol{W}_k \Vert_{*}+\frac{\eta}{2}\sum\nolimits_{k=1}^K\Vert \boldsymbol{W}_k \Vert_{F}^2}_{l_1\left(\boldsymbol{W}_k\right)}\nonumber\\
    &-\underbrace{\left(\rho\sum\nolimits_{k=1}^K\Vert \boldsymbol{W}_k \Vert_{2}+\frac{\eta}{2}\sum\nolimits_{k=1}^K\Vert \boldsymbol{W}_k \Vert_{F}^2\right)}_{l_2\left(\boldsymbol{W}_k\right)}.\tag{29}
\end{align}
The functions $l_1\left(\boldsymbol{W}_k\right)$ and $l_2\left(\boldsymbol{W}_k\right)$ are all $\eta$-strongly convex functions due to the introduction of quadratic terms $\frac{\eta}{2}\sum\nolimits_{k=1}^K\Vert \boldsymbol{W}_k \Vert_{F}^2$. By using the Fenchel's duality and SCA~\cite{fu2021reconfigurable,an2005dc}, we can obtain $\boldsymbol{W}_k^{(t)}$ at the $t$-th iteration by solving the following convex optimization problem:
\begin{subequations}
\begin{align}
{\bf (P5) }:\underset{\left\{\boldsymbol{W}_k,\mu_k\right\}_{k=1}^{K}}{\min}\quad& l_1\left(\boldsymbol{W}_k\right)-\sum_{k=1}^K\left \langle\boldsymbol{W}_k, \partial_{\boldsymbol{W}_k^{(t-1)}}l_2\left(\boldsymbol{W}_k\right) \right \rangle\nonumber \\
\text{s.t.}\quad\
&\text{(\ref{mu_k})},\text{(\ref{p3-1})},\text{(\ref{cons2_trans})},\ \text{and} \ \text{(\ref{semide_cons})}.\nonumber
\end{align}
\end{subequations}
Here, $\left \langle \boldsymbol{A},\boldsymbol{B}\right \rangle$ denotes the inner product of matrix $\boldsymbol{A}$ and $\boldsymbol{B}$, given by $\left \langle \boldsymbol{A},\boldsymbol{B}\right \rangle=\text{Re}\left(\text{tr}(\boldsymbol{A}^H\boldsymbol{B})\right)$. The sub-gradient of the function $l_2\left(\boldsymbol{W}_k\right)$ at point $\boldsymbol{W}_k^{(t-1)}$ can be calculated as
\begin{align}
    \partial_{\boldsymbol{W}_k^{(t-1)}}l_2\left(\boldsymbol{W}_k\right)=\rho \partial_{\boldsymbol{W}_k^{(t-1)}}\Vert \boldsymbol{W}_k\Vert_2+\eta \boldsymbol{W}_k^{(t-1)}.\tag{30}
\end{align}
$\partial_{\boldsymbol{W}_k^{(t-1)}}\Vert \boldsymbol{W}_k\Vert_2$ is further computed as
$\zeta_1^{(t-1)}(\zeta_1^{(t-1)})^H$, and $\zeta_1^{(t-1)}$ denotes the eigenvector corresponding to the largest eigenvalue of the matrix $\boldsymbol{W}_k^{(t-1)}$. Thus, the problem ${\bf (P5)}$ is recast as
\begin{subequations}
\begin{align}
{\bf (P6) }:&\nonumber\\
\underset{\left\{\boldsymbol{W}_k,\mu_k\right\}_{k=1}^{K}}{\min}&\quad l_1\left(\boldsymbol{W}_k\right)\!-\!\sum\nolimits_{k=1}^K\text{Re}\Bigg[\text{tr}\bigg(\boldsymbol{W}_k^H\nonumber \\
&\cdot \left(\rho\zeta_1^{(t-1)}(\zeta_1^{(t-1)})^H+\eta\boldsymbol{W}_k^{(t-1)}\right)\bigg)\Bigg]\nonumber \\
\text{s.t.}\quad\
&\begin{bmatrix}
   \bar{\boldsymbol{Q}}_k+\mu_k \boldsymbol{I}_M & \bar{\boldsymbol{r}}_k  \\
   \bar{\boldsymbol{r}}_k^H & \bar{{s}}_k-\mu_k d^2
  \end{bmatrix}\succeq 0,\forall k,\\
&\text{tr}\left( {{ {\left(\frac{N\boldsymbol{W}_k}{\bar{\gamma}_{c,k}}-\sum\nolimits_{i = 1,i \ne k}^K{\boldsymbol{W}_i}\right)\boldsymbol{C}_k }}}\right)\nonumber\\
&-\sum\nolimits_{i = 1,i \ne k}^K { \text{tr}\left(\boldsymbol{W}_i\boldsymbol{f}_k^H\boldsymbol{f}_k \right)}- \sigma _k^2\geq 0,\forall k,\\
&\boldsymbol{W}_k \succeq 0,\forall k,\\
&\mu_k\geq 0,\forall k.
\end{align}
\end{subequations}
The problem ${\bf (P6)}$ is convex and can be solved directly with CVX. We solve the problem ${\bf (P6)}$ iteratively until $\Vert \boldsymbol{W}_k \Vert_{*}-\Vert \boldsymbol{W}_k \Vert_{2}=0$. The overall algorithm of solving problem ${\bf (P1)}$ is summarized in Algorithm~\ref{Algorithm1_proposed}. 
\begin{remark}
    From the problem ${\bf (P6)}$, we observe that the transmit beamforming is optimized based on only the direct links $\{\boldsymbol{f}_k\}_{k=1}^{K}$ and reflective link covariance matrices $\{\boldsymbol{C}_k\}_{k=1}^{K}$, while $\boldsymbol{C}_k$ is calculated with the instantaneous CSIs of the reflective links $\left\{\boldsymbol{h}_l\right\}_{l=1}^L$ and $\left\{g_{l,k}\right\}_{k=1,l=1}^{K,L}$. We can utilize the channel correlations between the direct link and reflective links to avoid the use of instantaneous CSIs of the reflective links, and this approach is shown in the next section in detail.
\end{remark}

\begin{algorithm}[t!]
\caption{ The proposed algorithm to solve ${\bf (P1)}$}\label{Algorithm1_proposed}
\begin{algorithmic}[1]
\STATE \textbf{Input:} $\left\{\boldsymbol{f}_k\right\}_{k=1}^K$, $\left\{\boldsymbol{h}_l\right\}_{l=1}^L$, $\left\{g_{l,k}\right\}_{k=1,l=1}^{K,L}$, $\bar{{R}}_{s,k}$, $\bar{{R}}_{c,k}$, and $P_{out}$.
\STATE Calculate $\boldsymbol{C}_k$ with $\left\{\boldsymbol{h}_l\right\}_{l=1}^L$ and $\left\{g_{l,k}\right\}_{k=1,l=1}^{K,L}$.
\STATE  Drop the rank constraint (\ref{rank_cons}) and solve the rank-relaxed problem ${\bf (P3)}$. 
\STATE Denote the obtained solution as $\boldsymbol{W}_k^{\star}$.
\IF {the rank of $\boldsymbol{W}_k^{\star}$ is one}
\STATE Perform eigenvalue decomposition for $\boldsymbol{W}_k^{\star}$ as $\boldsymbol{W}_k^{\star}=\boldsymbol{U}_k^H\Sigma_k\boldsymbol{U}_k$. 
\STATE Obtain the optimal beamforming vector as $\boldsymbol{w}_k^{\star}=\boldsymbol{U}_k^H\boldsymbol{\Sigma}_k^{\frac{1}{2}}$.
\ELSE 
\STATE Let $t=1$, $\boldsymbol{W}_k^{(1)}=\boldsymbol{W}_k^{\star}$.
\WHILE{$\sum\nolimits_{k=1}^K\left(\Vert \boldsymbol{W}_k^{(t)} \Vert_{*}-\Vert \boldsymbol{W}_k^{(t)} \Vert_{2}\right) \neq 0$}
\STATE Compute the eigenvector corresponding to the largest eigenvalue of the matrix $\boldsymbol{W}_k^{(t)}$.
\STATE Solve the problem ${\bf (P6)}$ and denotes the obtained solution as $\boldsymbol{W}_k^{(t+1)}$.
\STATE Update $t=t+1$.
\ENDWHILE
\STATE Perform eigenvalue decomposition for $\boldsymbol{W}_k^{(t)}$ and obtain the optimal beamforming vector.
\ENDIF
\STATE \textbf{Output:} $\boldsymbol{w}_k^{\star}$ and the corresponding optimal objective function value $\sum\nolimits_{k=1}^K\Vert \boldsymbol{w}_k^{\star}\Vert^2$.
\end{algorithmic}
\end{algorithm}

\subsection{Complexity Analysis}
In the following, we investigate the computational complexity of the proposed algorithm. Recall that the rank-relaxed problem ${\bf (P3)}$ involves $K$ linear matrix inequality (LMI) constraints of size $M+1$, $K$ LMI constraints of size $M$, and $2K$ LMI constraints of size $1$. As such, it can be solved efficiently by a standard interior-point method with the total arithmetic cost less than~\cite{wang2014outage}:
\begin{align}
    &C^{(P3)}=\sqrt{2KM+3K}\cdot n\cdot \bigg[K\left((M+1)^3+M^3+2\right)\nonumber \\
    &+nK\left((M+1)^2+M^2+2\right)+n^2\bigg]\cdot \text{ln}\left(\frac{1}{\epsilon_s}\right),
\end{align}
where $n=\mathcal{O}\left(KM^2\right)$ and $\epsilon_s$ denotes the predefined accuracy of the solution. Note that the constraints in problem ${\bf (P6)}$ are the same as those in the rank-relaxed problem ${\bf (P3)}$, and thus the computational complexity is the same. Denote the iteration number for the SCA process by $I_{s}$ ($I_{s}=0$ if the rank of $\boldsymbol{W}_k^{\star}$ is one and, in this case, the SCA process is not applied). The computational complexity of the SCA process can be given by $I_{s}C^{(P3)}$. The overall computational complexity of the proposed algorithm is $\left(1+I_{s}\right)C^{(P3)}$.

\section{Special Case: A DoA-based Approach}\label{sec_specific}
In previous sections, we investigated a general multi-user multi-IoT-device symbiotic radio system and studied the transmit beamforming design for such a system. In this section, we consider a special case of the multi-user multi-IoT-device SR system where each cellular user is associated with several adjacent IoT devices, and multiple clusters are formed. A practical scenario for such a special case is the body area network, in which cellular users carry multiple wearable devices that need to transmit monitored data such as blood pressure and heart rate to the cellular user. In the considered special case, the user is located close to the associated IoT devices, and thus the direction-of-arrivals (DoAs) of the cellular user and the associated IoT device are highly correlated. Therefore, we propose a DoA-based approach that utilizes the DoA and angular spread (AS) of the direct links to design the transmit beamforming at the BS without the need for the instantaneous CSIs of the reflective links. Moreover, the transmit beamforming scheme proposed in previous sections is named as CSI-based approach. 

\begin{figure}[!t]
\centering
\includegraphics[width=\columnwidth] {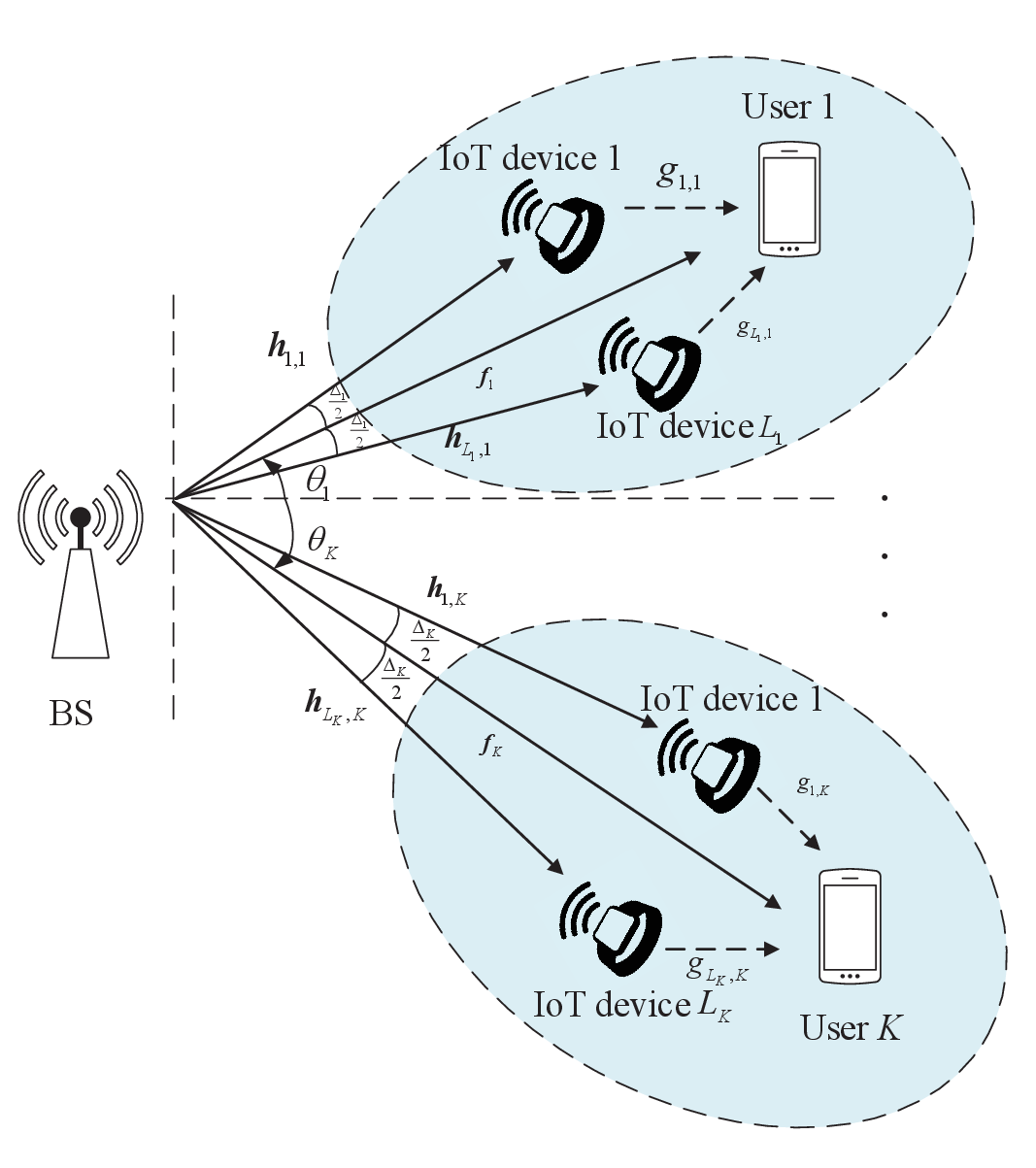}\vspace{-0.2cm}
\caption{The system model for the special case of the multi-user multi-IoT-device SR system.}
\label{fig:Fig_sim}
\vspace{-0.4cm}
\end{figure}

The system model for the special case is depicted in Fig.~\ref{fig:Fig_sim}. As shown in Fig.~\ref{fig:Fig_sim}, there are $L_k$ IoT devices associated with user $k$, and the links from the unassociated IoT devices to the cellular user are ignored due to relative weakness. Other setups are the same as the previous system model in Section II. In the following, we present the channel model, signal model, reflective link covariance matrix approximation, and transmit beamforming design for the special case in detail.  
\subsection{Channel Model}
The channel $\boldsymbol{f}_k$ from BS to user $k$ is modeled as a LoS channel, given by
\begin{align}
    \boldsymbol{f}_k=\beta_k \left(\theta_k\right)\boldsymbol{a}\left(\theta_k\right).
\end{align}
Here, $\beta_k\left(\theta_k\right)$ denotes the complex path strength of the signal coming from the DoA $\theta_k$, and $\boldsymbol{a}\left(\theta_k\right)$ denotes the array response vector for signals arriving from the DoA of $\theta_k$. By assuming uniform linear array at BS, $\boldsymbol{a}\left(\theta\right)$ is given by
\begin{align}
    \boldsymbol{a}\left(\theta\right)=[1,e^{j\chi\text{sin}(\theta)},\cdots,e^{j(M-1)\chi\text{sin}(\theta)}],
\end{align}
where $\chi=2\pi d/\lambda$, $\lambda$ is the wavelength of the carrier wave, and $d$ is the antenna spacing. Without loss of generality, we set $d=\frac{\lambda}{2}$. Similarly, the channel $\boldsymbol{h}_{l,k}$ from BS to IoT device $l$ associated with user $k$ is modeled as
\begin{align}
 \boldsymbol{h}_{l,k}=\beta_{l,k}\left(\theta_{l,k}\right) \boldsymbol{a}\left(\theta_{l,k}\right),\label{h_lk}
\end{align}
where $\beta_{l,k}\left(\theta_{l,k}\right)$ denotes the complex path strength of the signal coming from the DoA $\theta_{l,k}$. Since the user is located close to the associated IoT devices, in the following, we assume that all DoAs $\theta_{l,k}$ are uniformly distributed over $[\theta_k-\frac{\Delta_k}{2},\theta_k+\frac{\Delta_k}{2}]$, where $\Delta_k$ is the corresponding AS around $\theta_k$~\cite{xie2018channel}. The channel from the IoT device $l$ connected to user $k$ to the associated user, i.e., $g_{l,k}$ is an LoS channel with large-scale path loss, i.e., $g_{l,k}=\beta '_{l,k}$.

\subsection{Signal Model}
Use $k$ receives the direct link signal from the BS as well as the reflective link signals from the associated IoT devices. The received signal at user $k$ is given by
\begin{align}
{\tilde{y}_k}(n) = \boldsymbol{f}_k\boldsymbol{s}(n) + \sum\nolimits_{l = 1}^{L_k} {\alpha{g_{l,k}}{c_{l,k}}{\boldsymbol{h}_{l,k}}\boldsymbol{s}(n)}  + {u_k}(n),
\end{align}
where $\boldsymbol{h}_{l,k}$ denotes the channel from the BS to the IoT device $l$ associated to user $k$, and $c_{l,k}$ is the transmit symbol of IoT device $l$ associated with user $k$. User $k$ first decodes its information $s_k(n)$, and the corresponding cellular rate is expressed as
\begin{align}
    &{\tilde{R}_{s,k}} \nonumber\\
    &\!=\! {{{\log }_2}\left( {1 \!+\! \frac{{{{\left| {\left( {{\boldsymbol{f}_k} + \sum\nolimits_{l = 1}^{L_k} {\alpha_l {g_{l,k}}{c_{l,k}}{\boldsymbol{h}_{l,k}} } }\right){\boldsymbol{w}_k}} \right|}^2}}}{{\sum\nolimits_{i = 1,i \ne k}^K {{{\left| {\left( {{\boldsymbol{f}_k} + \sum\nolimits_{l = 1}^{L_k} {\alpha_l {g_{l,k}}{c_{l,k}}{\boldsymbol{h}_{l,k}}} } \right){\boldsymbol{w}_i}} \right|}^2}} \! +\! \sigma _k^2}}} \right)}.
\end{align}
After decoding its information, user $k$ decodes the information from the associated IoT devices by using SIC technique. The sum IoT device rate for user $k$ is given by (\ref{rck}), shown at the bottom of the next page.

\begin{figure*}[b]
    \begin{center}
    \begin{align}
{\tilde{R} _{c,k}} = \frac{1}{N}{\log_2}\left( {1 +\frac{N\sum\nolimits_{l = 1}^{L_k} {{{\left| {\alpha {g_{l,k}}{\boldsymbol{h}_l}{\boldsymbol{w}_k}} \right|}^2}}}{{\sum\nolimits_{i = 1,i \ne k}^K {{{\left| {{\boldsymbol{f}_k}{\boldsymbol{w}_i}} \right|}^2}}  + \sum\nolimits_{l = 1}^{L_k} {\sum\nolimits_{i = 1,i \ne k}^K {{{\left| {\alpha {g_{l,k}}{\boldsymbol{h}_l}{\boldsymbol{w}_i}} \right|}^2}} }  + \sigma _k^2}}}\right).\label{rck}
\end{align}
\end{center}
\end{figure*}

\subsection{Reflective Link Covariance Matrix Approximation}

Denote  $\Delta\boldsymbol{f}_{k}=\sum\nolimits_{l = 1}^{L_k} {\alpha {g_{l,k}}{c_{l,k}}{\boldsymbol{h}_{l,k}} } $ as the reflective link of user $k$, which follows the distribution of $\mathcal{CN}(\boldsymbol{0},\boldsymbol{C}_k)$. The covariance matrix $\boldsymbol{C}_k$ is given by 
\begin{align}
   \boldsymbol{C}_k&=\sum\nolimits_{l = 1}^{L_k} {|\alpha {g_{l,k}}{|^2}\boldsymbol{h}_{l,k}^H{\boldsymbol{h}_{l,k}}}\nonumber\\
   &=\sum\nolimits_{l = 1}^{L_k} {|\alpha {\beta '_{l,k}} \beta_{l,k}\left(\theta_{l,k}\right){|^2}\boldsymbol{a}^H\left(\theta_{l,k}\right)\boldsymbol{a}\left(\theta_{l,k}\right)}. \label{C_k}
\end{align}
When $L_k$ is large and suppose that $\theta_{l,k}$s are dense in the range $[\theta_k-\frac{\Delta_k}{2},\theta_k+\frac{\Delta_k}{2}]$~\cite{liang2001downlink}, we can rewrite $\boldsymbol{C}_k$ as~\footnote{Note that when $\Delta_k$ is small, $L_k$ does not need to be large to achieve a very accurate approximation. For example, when $\Delta_k$ takes the value of 0.01, which is a typical value in the body area network scenario, $L_k=4$ can indeed achieve a good approximation, as shown in the simulations. }
\begin{align}
    \boldsymbol{C}_k\approx\frac{L_k}{\Delta_k}\int_{\theta_k-\frac{\Delta_k}{2}}^{\theta_k+\frac{\Delta_k}{2}}|\alpha {\beta '_{l,k}} \beta_{l,k}\left(\theta_{l,k}\right){|^2}\boldsymbol{a}^H\left(\theta\right)\boldsymbol{a}\left(\theta\right)d \theta.
\end{align}
Since the IoT devices are located close to the cellular user, we further assume that the large-scale path loss of the reflective link with the same user is the same, i.e.,  $\psi_k\triangleq \beta '_{l,k}\beta_{l,k}\left(\theta_{l,k}\right)$ is a constant and independent of $\theta_{l,k}$. We can further obtain
\begin{align}
    \boldsymbol{C}_k=\frac{|\alpha \psi_k {|^2}L_k}{\Delta_k}\int_{\theta_k-\frac{\Delta_k}{2}}^{\theta_k+\frac{\Delta_k}{2}}\boldsymbol{a}^H\left(\theta\right)\boldsymbol{a}\left(\theta\right)d \theta.
\end{align}
 The $(m,n)$-th element of $\boldsymbol{C}_k$ is expressed as
\begin{align}
    &\left[\boldsymbol{C}_k\right]_{m,n}\nonumber\\
    &=\frac{|\alpha \psi_k {|^2}L_k}{\Delta_k}\int_{\theta_k-\frac{\Delta_k}{2}}^{\theta_k+\frac{\Delta_k}{2}}e ^{-j(m-n)\chi \text{sin}(\theta)}d \theta \nonumber\\
    &\approx \frac{|\alpha \psi_k {|^2}L_k}{\Delta_k}\int_{\theta_k-\frac{\Delta_k}{2}}^{\theta_k+\frac{\Delta_k}{2}}e ^{-j(m-n)\chi \left[\text{sin}(\theta_k)+(\theta-\theta_k)\text{cos}(\theta_k)\right]}d \theta \nonumber\\
    &=\frac{|\alpha \psi_k {|^2}L_k}{\Delta_k}e ^{-j(m-n)\chi \text{sin}(\theta_k)}\int_{-\frac{\Delta_k}{2}}^{\frac{\Delta_k}{2}}e ^{-j(m-n)\chi \text{cos}(\theta_k)\theta}d \theta\nonumber\\
    &=|\alpha \psi_k {|^2}L_ke ^{-j(m-n)\chi \text{sin}(\theta_k)}\text{sinc}\left(\frac{\Delta_k}{2}(m-n)\chi\text{cos}(\theta_k)\right). \label{Ck_final}
\end{align}
From (\ref{Ck_final}), we can find that the approximation of $\boldsymbol{C}_k$ is determined by the values of central DoA $\theta_k$, angle spread $\Delta_k$ and the large scale path loss $\psi_k$. Therefore, when designing the beamforming vectors, the BS does not require the instantaneous CSI but the information of $\theta_k$, $\Delta_k$, and $\psi_k$, leading to a significant reduction in channel feedback overhead.


\subsection{Transmit Beamforming Design: A DoA-based Approach}
For the considered special case, the detailed transmission protocol and transmit beamforming design method is described as follows. The cellular users first perform channel estimation to obtain the direct link channels with the BS's pilot symbols. The corresponding DoAs and ASs of the direct links can then be estimated with classic MUSIC algorithms~\cite{chung2014doa,jeong2002performance,jeong2001generalization}. The users then feedback the direct link channels, DoAs, and ASs to the BS. The BS can then calculate the approximation of the reflective link covariance matrix with the feedback DoAs and ASs by using (\ref{Ck_final}). The transmit beamforming at the BS can then be optimized with algorithm 1 by replacing the second step with the approximated $\boldsymbol{C}_k$.

\section{Numerical Results}\label{sec-simulation}
\begin{figure}[!t]
\centering
\includegraphics[width=\columnwidth] {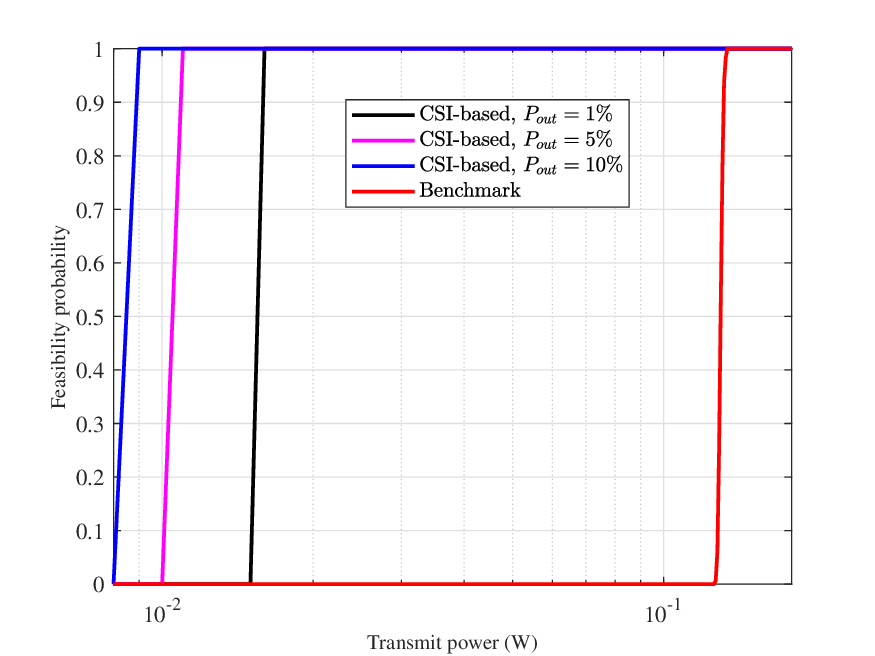}\vspace{-0.2cm}
\caption{Feasibility versus the transmit power with different outage requirement $P_{out}$ when $\bar{{R}}_{s,k}=3$ bps/Hz and $\bar{{R}}_{c,k}=0.12$ bps/Hz.}
\label{fig:Fig_feasibility}
\vspace{-0.5cm}
\end{figure}

In this section, numerical results are provided to substantiate the performance of the multi-user multi-IoT-device SR system and the proposed algorithm.  
The BS is equipped with six transmit antennas and serves two users, while each user is associated with four IoT devices, i.e., $M=6$, $K=2$, and $L_1=L_2=\bar{L}=4$. The noise power at user $k$ is set to $\sigma_k^2=\sigma^2=-100 \ \text{dBm}, \forall k$. The reflection coefficient is set to $\alpha=0.5$. The symbol period of IoT transmission is $16$ times that of the BS, i.e., $N=16$. The distances between BS and two users are $d_1=200$ m, $d_2=180$ m. We adopt the path loss model in~\cite{long2019symbiotic,guo2018exploiting}, and the path losses for the direct link channels are calculated as follows:
\begin{align}
    PL_k=\frac{\lambda_c^2G_bG_r}{(4\pi)^2d_k^{\nu}}, k=1,2,
\end{align}
where $\lambda_c$ denoting the wavelength of the carrier signal is set to 0.33 m, $G_b$, $G_r$ denoting the antenna gains of BS and user are set to $G_b=G_r=6$ dB, $\nu$ denoting the path loss exponent is set to $\nu=3.5$. The strength of the reflective links is 20 dB weaker than those of the direct links, i.e., $| \psi_k {|^2}=10^{-2} PL_k$. The DoAs are set to $-\frac{\pi}{3}$ and $\frac{\pi}{3}$. Unless otherwise specified, the angular spread is set to 0.01 for both users. The penalty ratio is set to $\rho=10$ and $\eta=1$. 

For comparison, we consider the maximum ratio transmission (MRT) beamforming scheme as the benchmark scheme~\cite{zhang2016mrt}. In Fig.~\ref{fig:Fig_feasibility}, we plot the system feasibility probability with respect to the transmit power. Here, feasibility probability denotes the probability that all the cellular transmission outage requirements and IoT transmission sum rate requirements are satisfied. It is observed that the performance of the proposed algorithm outperforms that of the benchmark scheme since a higher feasibility probability can be achieved with a lower transmit power. Even when the outage probability is stringent, i.e., $P_{out}=1\%$, the proposed scheme can still achieve a relatively good performance.

\begin{figure}[!t]
\centering
\includegraphics[width=\columnwidth] {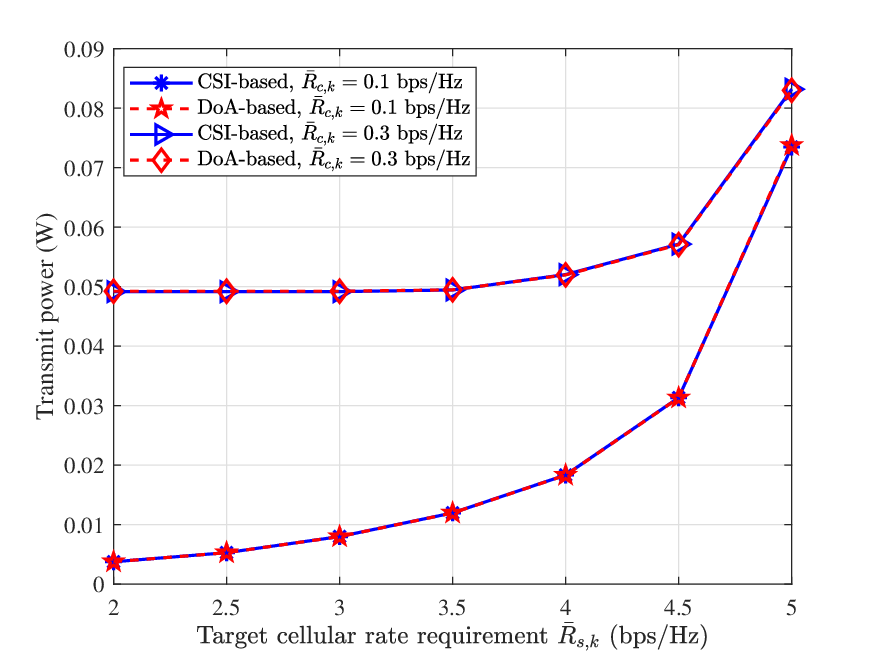}\vspace{-0.2cm}
\caption{Transmit power versus the target cellular rate requirement $\bar{{R}}_{s,k}$ with different $\bar{{R}}_{c,k}$.}
\label{fig:Fig_simu_1}
\vspace{-0.5cm}
\end{figure}

Fig.~\ref{fig:Fig_simu_1} shows the transmit power versus target cellular rate requirement $\bar{{R}}_{s,k}$ with different $\bar{{R}}_{c,k}$ when the outage probability is set to $P_{out}=10\%$. It is first observed that the CSI-based and DoA-based curves show the same transmit power performance, which verifies the effectiveness of using only the DoA and AS for transmit beamforming design. We then observe that the required transmit power increases as $\bar{{R}}_{s,k}$ increases. Besides, a higher $\bar{{R}}_{c,k}$ leads to a higher transmit power. When $\bar{{R}}_{s,k}$ is relatively low, the difference in transmit power between different $\bar{{R}}_{c,k}$ is significant. However, as $\bar{{R}}_{s,k}$ increases, the impact of $\bar{{R}}_{c,k}$ on the transmit power decreases. For example, when $\bar{{R}}_{s,k}$ takes the value of 2 bps/Hz, the difference of transmit power is 0.0455 W when $\bar{{R}}_{c,k}$ takes the value of 0.1 bps/Hz and 0.3 bps/Hz. When $\bar{{R}}_{s,k}$ is 5 bps/Hz, the difference of transmit power decreases to 0.0093 W. The reason is that when $\bar{{R}}_{s,k}$ is low, the transmit power is restricted by the IoT transmission sum rate constraint and thus increasing $\bar{{R}}_{c,k}$ has a significant impact on the transmit power. On the other hand, when $\bar{{R}}_{s,k}$ is high, and in this case, the cellular outage constraint is stringent, increasing $\bar{{R}}_{c,k}$ does not have a noticeable effect on the transmit power.

\begin{figure}[!t]
\centering
\includegraphics[width=\columnwidth] {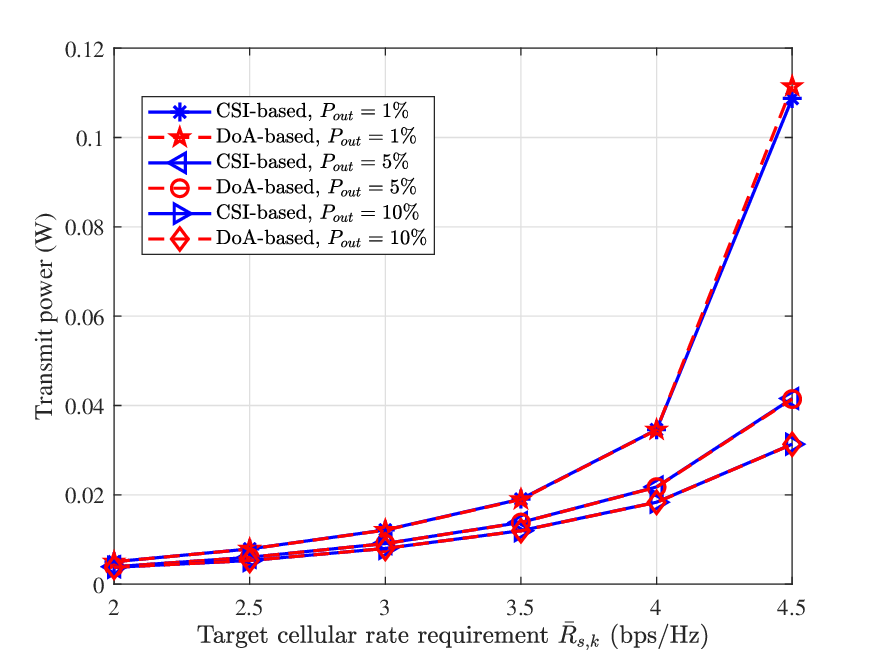}\vspace{-0.2cm}
\caption{Transmit power versus the target cellular rate requirement $\bar{{R}}_{s,k}$ with different outage requirement $P_{out}$.}
\label{fig:Fig_simu_2}
\vspace{-0.5cm}
\end{figure}

In Fig.~\ref{fig:Fig_simu_2}, we show the transmit power with respect to the target cellular rate requirement $\bar{{R}}_{s,k}$ with different outage requirement $P_{out}$ when $\bar{{R}}_{c,k}$ is set to 0.1 bps/Hz. From Fig.~\ref{fig:Fig_simu_2}, we observe that the DoA-based method yields performance equivalent to the CSI-based method, irrespective of the value of $P_{out}$. A lower outage probability $P_{out}$ represents a more stringent outage requirement for cellular transmission. The lower $P_{out}$ is, the higher the transmit power is. Moreover, when $\bar{{R}}_{s,k}$ is low, the transmit power does not show a significant difference for different outage requirements $P_{out}$ since the cellular outage constraints are easy to be met in this case. When $\bar{{R}}_{s,k}$ is high, the transmit power shows an obvious difference when $P_{out}$ takes different values. As Fig.~\ref{fig:Fig_simu_2} shows, the transmit power reaches 0.11 W when $P_{out}=1\%$ while the transmit power is 0.041 W and 0.031 W when $P_{out}$ is set to $5\%$ and $10\%$, respectively.

\begin{figure}[!t]
\centering
\includegraphics[width=\columnwidth] {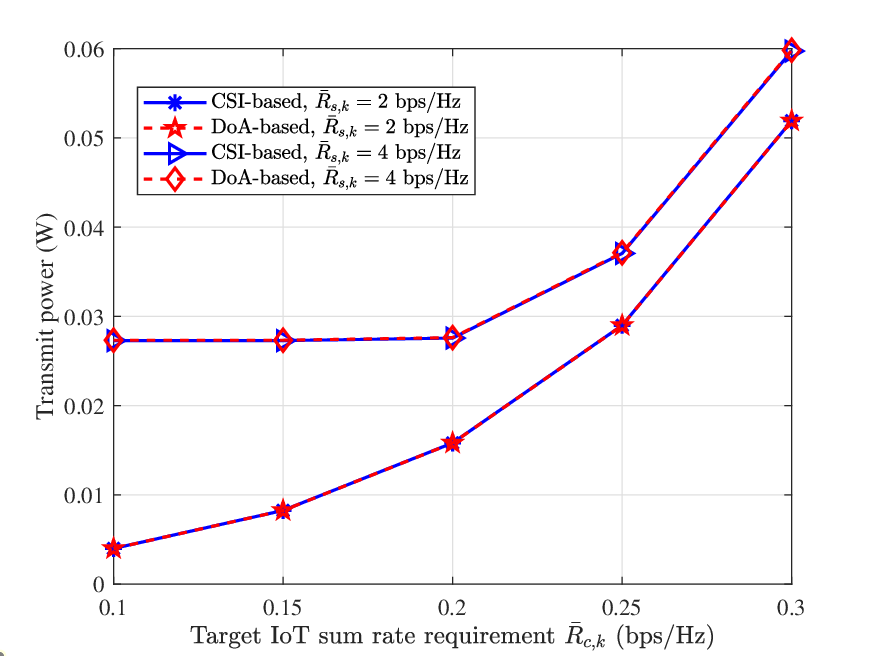}\vspace{-0.2cm}
\caption{Transmit power versus the target IoT sum rate requirement $\bar{{R}}_{c,k}$ with different $\bar{{R}}_{s,k}$.}
\label{fig:Fig_simu_3}
\vspace{-0.5cm}
\end{figure}

Fig.~\ref{fig:Fig_simu_3} shows the transmit power versus target IoT sum rate requirement $\bar{{R}}_{c,k}$ with different $\bar{{R}}_{s,k}$ when the outage requirement $P_{out}$ is set to $10\%$. The curves in Fig.~\ref{fig:Fig_simu_3} show a similar trend to those in Fig.~\ref{fig:Fig_simu_1}. Specifically, as seen in Fig.~\ref{fig:Fig_simu_3}, the transmit power increases with the increase of $\bar{{R}}_{c,k}$ when $\bar{{R}}_{s,k}=2$ bps/Hz, while when $\bar{{R}}_{s,k}=4$ bps/Hz it remains unchanged when $\bar{{R}}_{c,k}$ is lower than 0.2 bps/Hz and then increases as $\bar{{R}}_{c,k}$ increases. When $\bar{{R}}_{c,k}$ is high, the difference of transmit power caused by increasing $\bar{{R}}_{s,k}$ will be negligible.

\begin{figure}[!t]
\centering
\includegraphics[width=\columnwidth] {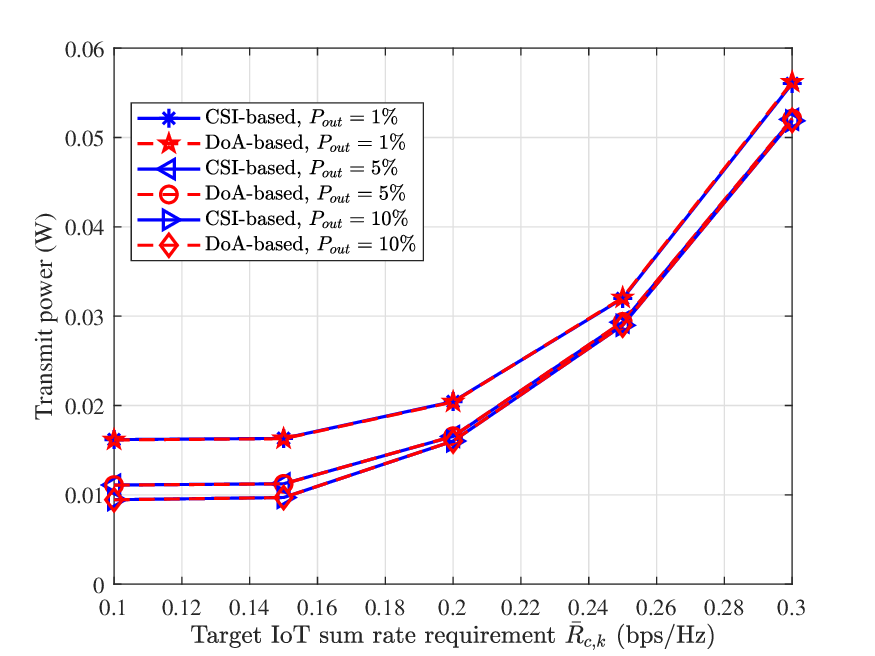}\vspace{-0.2cm}
\caption{Transmit power versus the target IoT sum rate requirement $\bar{{R}}_{c,k}$ with different outage requirement $P_{out}$.}
\label{fig:Fig_simu_4}
\vspace{-0.5cm}
\end{figure}
Fig.~\ref{fig:Fig_simu_4} illustrates the transmit power versus target IoT sum rate requirement $\bar{{R}}_{c,k}$ with different outage requirement $P_{out}$ when $\bar{{R}}_{s,k}$ is set to 3 bps/Hz. We first focus on the curves when $P_{out}$ takes the values of $10\%$ and $5\%$. We observe that there is a minor discrepancy in the performance of these curves when $\bar{{R}}_{c,k}$ is low. However, with an increase in $\bar{{R}}_{c,k}$, the performance of these curves gradually approaches and overlaps. This phenomenon can be attributed to the fact that $P_{out}$ only affects the cellular transmission performance, whereas the transmit power is more constricted by the IoT transmission as $\bar{{R}}_{c,k}$ increases. When a stricter outage probability requirement of $1\%$ is considered, the difference in transmission power is considerable. Furthermore, even for larger values of $\bar{{R}}_{c,k}$, such as 0.3 bps/Hz, a significant difference in transmit power is still observed.

\begin{figure}[!t]
\centering
\includegraphics[width=\columnwidth] {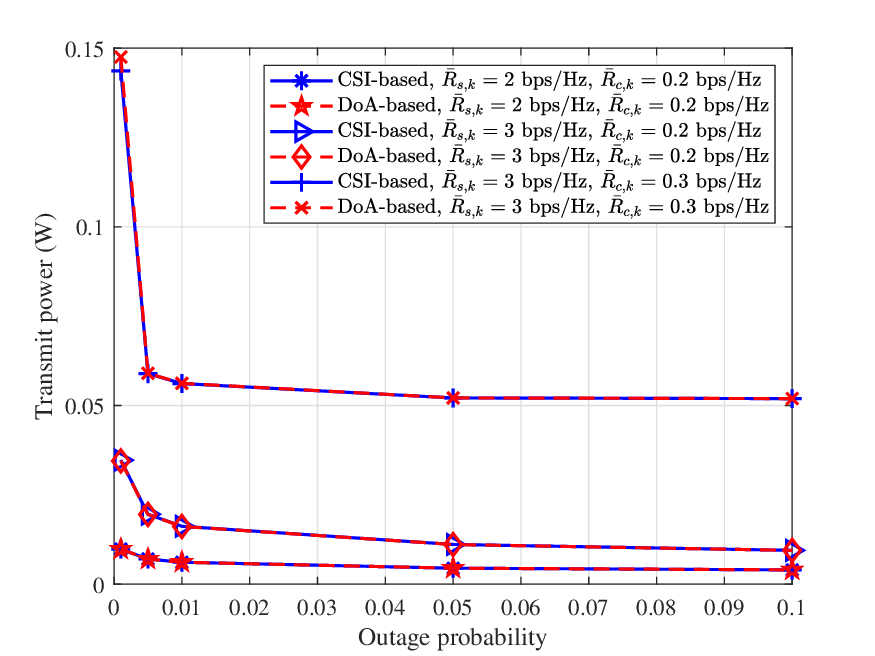}\vspace{-0.2cm}
\caption{Transmit power versus the outage requirement $P_{out}$.}
\label{fig:Fig_simu_5}
\vspace{-0.5cm}
\end{figure}
Fig~\ref{fig:Fig_simu_5} shows the transmit power versus outage requirement $P_{out}$. We can observe that as the outage probability increases, the transmit power decreases significantly and gradually approaches a constant value. Moreover, the curves in Fig~\ref{fig:Fig_simu_5} are consistent with their counterparts in Fig.~\ref{fig:Fig_simu_2} and Fig.~\ref{fig:Fig_simu_4}.

\begin{figure}[!t]
\centering
\includegraphics[width=\columnwidth] {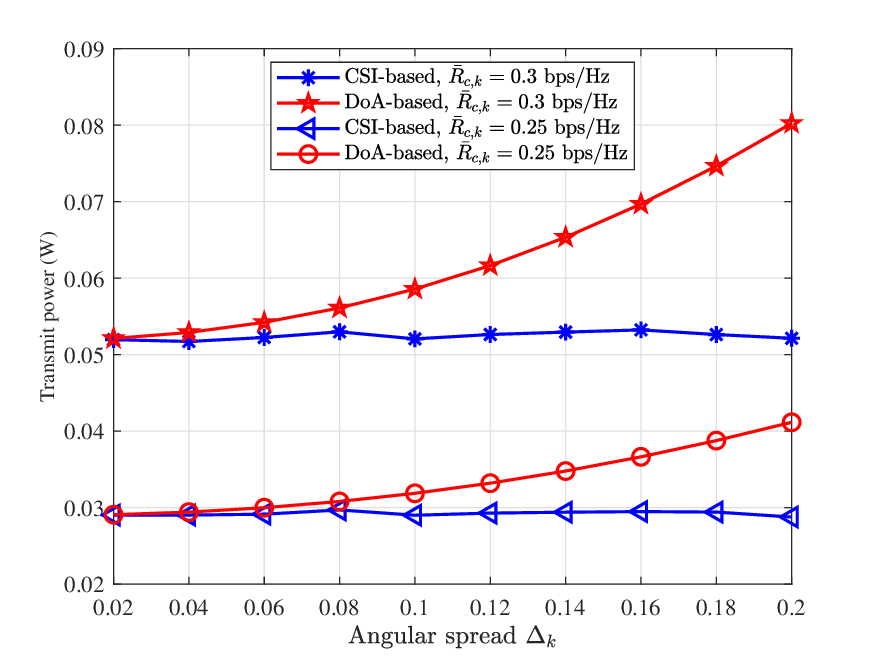}\vspace{-0.2cm}
\caption{Transmit power versus the angular spread $\Delta_k$.}
\label{fig:Fig_simu_AS}
\vspace{-0.5cm}
\end{figure}

In Fig.~\ref{fig:Fig_simu_AS}, we investigate the impact of angular spread on transmit power. The angular spread refers to the extent of the distribution of reflective devices in the vicinity of the user, with larger values indicating a wider coverage area. It can be observed that as the angular spread decreases from 0.2 to 0.02, the DoA-based approach gradually achieves comparable performance to the CSI-based approach in terms of transmit power. This can be attributed to the fact that a smaller angular spread allows for a more accurate approximation of the reflective link covariance matrix, whereas a larger angular spread results in a less accurate approximation. Moreover, when $\bar{{R}}_{c,k}=0.25$ bps/Hz, the DoA-based approach can achieve a similar performance as the CSI-based when the angular spread is less than 0.06.
Furthermore, the CSI-based approach exhibits consistent transmit power performance regardless of the angular spread.

\begin{figure}[!t]
\centering
\includegraphics[width=\columnwidth] {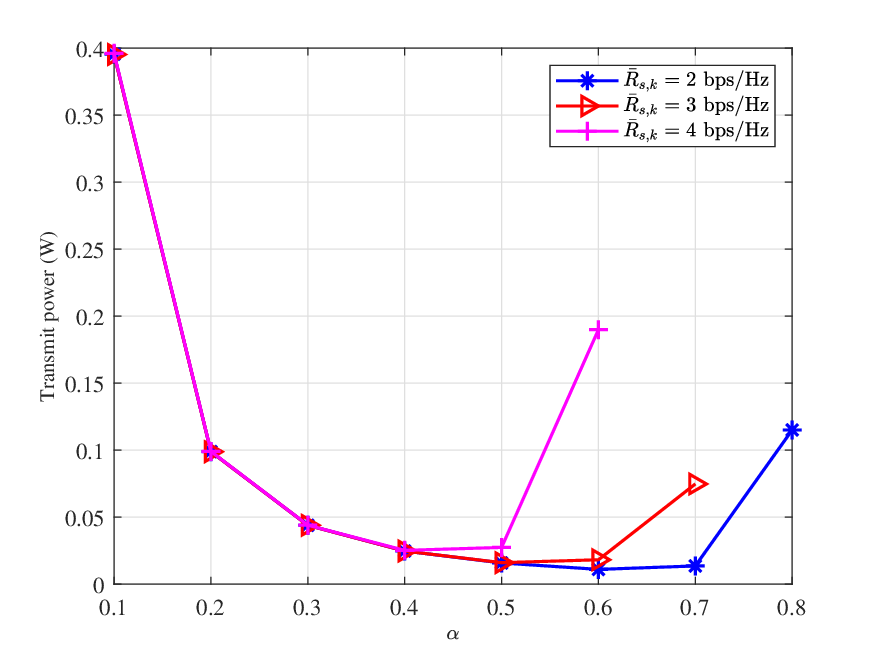}\vspace{-0.2cm}
\caption{Transmit power versus the reflection coefficient $\alpha$.}
\label{fig:Fig_simu_6}
\vspace{-0.5cm}
\end{figure}
Fig.~\ref{fig:Fig_simu_6} shows the transmit power versus reflection coefficient $\alpha$ with different $\bar{{R}}_{s,k}$. $\bar{{R}}_{c,k}$ is set to 0.2 bps/Hz and $P_{out}$ is set to $10\%$. As the reflection coefficient $\alpha$ gradually increases from 0.1 to 0.4, the transmit power decreases continuously for different values of $\bar{{R}}_{s,k}$, and the corresponding curves overlap with each other, as shown in Fig.~\ref{fig:Fig_simu_6}. However, further increasing the reflection coefficient $\alpha$ leads to an increasing trend in the required transmit power, and a higher $\bar{{R}}_{s,k}$ corresponds to a lower reflection coefficient that results in the lowest transmit power. There exists an optimal reflection coefficient that can minimize the transmit power for different $\bar{{R}}_{s,k}$. The reason is that when the reflection coefficient is small, the reflection paths are relatively weak, and the IoT transmission rate requirement is the major constraint that restricts the transmit power. With the increase of $\alpha$, the reflection paths get stronger, and the transmit power decreases. On the other hand, the reflection link is treated as a channel uncertainty of the direct link, and according to (\ref{Ck_final}), the larger the reflection coefficient, the greater the channel uncertainty. When $\alpha$ is larger than a threshold, for example, 0.4 when $\bar{{R}}_{s,k}=4$ bps/Hz, the cellular transmission outage constraint becomes the one that restricts the transmit power. Therefore, as $\alpha$ increases, the transmit power increases as well.

\begin{figure}[!t]
\centering
\includegraphics[width=\columnwidth] {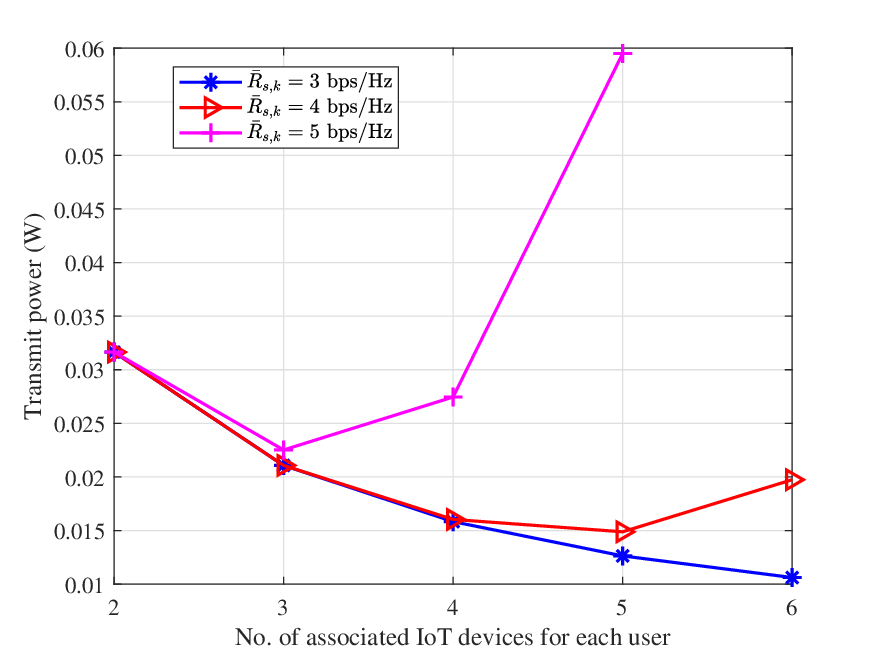}\vspace{-0.2cm}
\caption{Transmit power versus the number of associated IoT devices for each user with different $\bar{{R}}_{s,k}$.}
\label{fig:Fig_simu_7}
\vspace{-0.5cm}
\end{figure}
Finally, we vary the number of associated IoT devices for each user $\bar{L}$ and plot Fig.~\ref{fig:Fig_simu_7}, where $\bar{{R}}_{c,k}$ is set to 0.2 bps/Hz and $\bar{{R}}_{c,k}$ is set to $10\%$. We can observe that when $\bar{{R}}_{c,k}$ takes the value of 3 bps/Hz, the transmit power decreases as $\bar{L}$ increases, while it first decreases and then increases when $\bar{{R}}_{c,k}$ equals 4 and 5 bps/Hz. The reason for this phenomenon is quite similar to Fig.~\ref{fig:Fig_simu_6}. The more IoT devices are associated with the user, the stronger the reflection paths are. When $\bar{L}$ is small, the IoT transmission rate constraint works, while the cellular transmission outage constraint works when $\bar{L}$ is large. There exists an optimal number of connected IoT devices with which the transmit power is minimized. This phenomenon gives us the intuition that some IoT devices can be silent to achieve power saving.

\section{Conclusions}\label{sec-conclusion}
 In this paper, we have proposed the multi-user multi-IoT-device SR system as a novel massive access scheme for cellular IoT, where multiple IoT devices simultaneously backscatter their information to multiple users over the cellular signal. In consideration of the channel uncertainty introduced by the IoT transmission, we have formulated a transmit beamforming design problem with the aim of minimizing the transmit power at the BS under the cellular transmission outage probability constraints and IoT transmission sum rate constraints. $\mathcal{S}$-lemma, SDP, and DC techniques are utilized to transform the non-convex original problem into a series of convex DC problems. We have also considered a special case of the multi-user multi-IoT-device SR and proposed a DoA-based transmit beamforming approach to reduce the channel feedback overhead. The DoA-based approach shows comparative performance with the CSI-based approach when the angular spreads are small. Extensive simulation results have verified the multi-user multi-IoT-device SR system and the effectiveness of the proposed algorithm.

\bibliography{reference}
\bibliographystyle{IEEEtran}
\end{document}